\documentclass[journal]{IEEEtran}

\usepackage{cite}
\usepackage{amsmath}
\usepackage{amssymb}
\usepackage{array}
\usepackage{dblfloatfix}
\usepackage{url}
\usepackage{enumitem}
\usepackage{bm}
\usepackage[normalem]{ulem}

\usepackage[pdftex]{graphicx}
\graphicspath{{./Figures/}}

\usepackage[framemethod=tikz]{mdframed}

\usepackage{booktabs}
\newcommand{\ra}[1]{\renewcommand{\arraystretch}{#1}}
\usepackage[tableposition=top]{caption}

\usepackage{color}
\usepackage{xcolor}
\usepackage{empheq}
\usepackage[most]{tcolorbox}
\definecolor{mygray}{gray}{0.45}
\newenvironment{graytext}{\color{mygray}}{\ignorespacesafterend}

\newcommand\numberthis{\addtocounter{equation}{1}\tag{\theequation}}

\begin{document}

\title{Unsupervised Low Latency Speech Enhancement\\with RT-GCC-NMF}

\author{Sean~U.~N.~Wood~and~Jean~Rouat,~\IEEEmembership{Senior Member,~IEEE}
\thanks{S. Wood and J. Rouat are affiliated with NECOTIS, Department of Electrical and Computer Engineering, University of Sherbrooke, Sherbrooke, QC, J1K 2R1, Canada. e-mail: \{sean.wood, jean.rouat\}@usherbrooke.ca\newline S. Wood is also affiliated with the Signal Processing and Speech Communication Lab, Graz University of Technology, Graz 8010, Austria}%
}

\markboth{IEEE~JOURNAL~OF~SELECTED~TOPICS~IN~SIGNAL~PROCESSING,~Vol.~X,~No.~X,~201X}%
{Shell \MakeLowercase{\textit{et al.}}: Bare Demo of IEEEtran.cls for IEEE Journals}

\maketitle

\IEEEpubid{\raisebox{-7mm}{
\begin{minipage}{\textwidth}\ \\[12pt] \centering
1932-4553 (c) 2018 IEEE. Personal use is permitted, but republication/redistribution requires IEEE permission.\\
See http://www.ieee.org/publications\_standards/publications/rights/index.html for more information.\\
Citation information: DOI 10.1109/JSTSP.2019.2909193, IEEE Journal of Selected Topics in Signal Processing
\end{minipage}}} 

\begin{abstract}
In this paper, we present RT-GCC-NMF: a real-time (RT), two-channel blind speech enhancement algorithm that combines the non-negative matrix factorization (NMF) dictionary learning algorithm with the generalized cross-correlation (GCC) spatial localization method.
Using a pre-learned universal NMF dictionary, RT-GCC-NMF operates in a frame-by-frame fashion by associating individual dictionary atoms to target speech or background interference based on their estimated time-delay of arrivals (TDOA).
We evaluate RT-GCC-NMF on two-channel mixtures of speech and real-world noise from the Signal Separation and Evaluation Campaign (SiSEC).
We demonstrate that this approach generalizes to new speakers, acoustic environments, and recording setups from very little training data, and outperforms all but one of the algorithms from the SiSEC challenge in terms of overall Perceptual Evaluation methods for Audio Source Separation (PEASS) scores and compares favourably to the ideal binary mask baseline.
Over a wide range of input SNRs, we show that this approach simultaneously improves the PEASS and signal to noise ratio (SNR)-based Blind Source Separation (BSS) Eval objective quality metrics as well as the short-time objective intelligibility (STOI) and extended STOI (ESTOI) objective speech intelligibility metrics.
A flexible, soft masking function in the space of NMF activation coefficients offers real-time control of the trade-off between interference suppression and target speaker fidelity.
Finally, we use an asymmetric short-time Fourier transform (STFT) to reduce the inherent algorithmic latency of RT-GCC-NMF from 64 ms to 2 ms with no loss in performance.
We demonstrate that latencies within the tolerable range for hearing aids are possible on current hardware platforms.
\end{abstract}


\section{Introduction}

\IEEEPARstart{R}{eal-world} speech processing applications including assistive listening devices and digital personal assistants rely on online speech enhancement algorithms to suppress noise and interfering sound sources.
However, a significant amount of research has focused on the offline setting with many algorithms being unsuitable for real-time use due to batch processing or computational requirements.
Recent speech enhancement and source separation approaches based on deep neural networks offer impressive performance gains compared with traditional real-time signal processing methods \cite{wang2018multi, wang2018integrating, williamson2017time, kolbk2017speech}, however these methods tend to be computationally demanding, preventing their use in low-power devices, and often rely on future information, preventing their use in real-time systems.
Deep learning methods also require a significant amount of supervised data for training, thus preventing their use in data-poor domains. 
The offline GCC-NMF source separation algorithm \cite{wood2017blind}, on the other hand, is an unsupervised approach that performs feature learning on the mixture signal itself, thus forgoing the need for large amounts of supervised training data.
GCC-NMF combines unsupervised machine learning via non-negative matrix factorization (NMF) \cite{lee2001algorithms} with the generalized cross-correlation (GCC) spatial localization method rooted in signal processing \cite{knapp1976generalized}.
However, the NMF dictionary, its activation coefficients, and the target speaker's time difference of arrival (TDOA) are all estimated using entire noisy utterances (10 seconds in duration), thus precluding its use in real-time.\\
\indent NMF learns parts-based representations from non-negative data \cite{lee1999learning}.
For audio signals, NMF is typically applied to magnitude spectrogram representations, learning spectral or spectro-temporal atoms that capture patterns typical of sound sources \cite{schmidt2006single}.
In the context of speech enhancement, we must then determine which atoms belong to the target speaker and which belong to interference.
Supervised model-based approaches solve this problem by pre-learning dictionaries for each source in isolation \cite{schmidt2006single, smaragdis2007supervised}, allowing for real-time operation as only the current (and possibly previous) spectrogram frames are required at runtime \cite{joder2012real}.
Unsupervised model-based approaches leverage the spatial distribution of the underlying sources to learn individual source dictionaries with no prior information in the form of separate datasets for speech and noise \cite{ozerov2012general}.
These unsupervised approaches are unable to operate in real-time as the spatial information does not generalize to unseen settings.\\
\indent In this work, we present RT-GCC-NMF\footnote{Source code is available at \url{https://www.github.com/seanwood/gcc-nmf}}: a low latency, two-channel speech enhancement algorithm that requires very little unsupervised training data and runs in real-time on low powered hardware platforms\footnote{Preliminary elements of this paper were presented in \cite{wood2017realtimegccnmf, wood2017gccnmfdemo, wood2017chat}}.
RT-GCC-NMF differs from previous NMF-based approaches by pre-learning a \emph{single} NMF dictionary on random single-channel speech and noise signals in a purely unsupervised fashion.
The dictionary learning algorithm therefore encodes a random mixture of speech and noise, with no knowledge of which features belong to which.
The resulting \emph{universal} model contains atoms useful for encoding both speech and noise components.
While no spatial information is encoded in the dictionary atoms themselves, we leverage the TDOA information from the two microphones at runtime to associate individual NMF dictionary atoms with the target or interference on frame-by-frame basis based on the dictionary atom's estimated TDOA.
The target estimate is reconstructed using only the atoms associated with the target, as is typical for NMF-based speech enhancement \cite{king2012new, mohammadiha2013supervised}.
This approach combines the flexibility of unsupervised NMF-based speech enhancement requiring no prior knowledge of differences between speech and noise characteristics, with online operation allowing for real-time use.
RT-GCC-NMF generalizes to unseen speakers, acoustic environments, and recording setups from very little unlabeled training data: on the order of one thousand 64 ms frames, compared to hours of labeled training data required for deep learning approaches \cite{williamson2017time}.
The pre-learned NMF dictionary is also very fast to train, on the order of seconds or minutes, in contrast with hours required to train deep neural networks.\\
\indent The STFT underlying most speech enhancement techniques based on deep neural networks or NMF, including the approach we present here, brings trade-off between spectral resolution and the inherent delay between the system's input and output.
This algorithmic latency is independent of processing speed, and is a consequence of the temporal windowing underlying the STFT.
Since many algorithms rely on high spectral resolution, algorithmic latencies greater than 64 ms are common.
For assistive listening devices including hearing aids, however, such high latencies lead to the perception of objectionable echoes as a superposition of both the aided and unaided sounds are heard by the listener \cite{stone1999tolerable}.
Depending on the type and severity of hearing loss, delays below 15 to 32 ms are likely required to be tolerable \cite{stone2005tolerable, herbig2010acceptable}, with delays less than 10 ms being a reasonable objective in the general case \cite{agnew2000just, dillon2003sound}.
To address this problem, an asymmetric STFT windowing approach proposed by Mauler and Martin \cite{mauler2007low} is combined with RT-GCC-NMF, simultaneously providing high spectral resolution and latencies well below the 10 ms target.\\
\indent The contributions of this paper are organized as follows.
In Section \ref{sec:offline-gcc-nmf}, we review GCC-NMF \cite{wood2017blind} and propose a flexible soft masking function in the space of NMF activation coefficients.
In Section \ref{subsec:experiments-window-parameters}, we show that this mask provides frame-by-frame control of the trade-off between target fidelity and interference suppression.
In Section \ref{sec:online-gcc-nmf}, we develop RT-GCC-NMF, pre-learning an NMF dictionary in an unsupervised fashion on a different dataset than used at test time.
In Section \ref{subsec:dictionary-pretraining}, we show that this approach generalizes to unseen speakers, acoustic conditions, and recording setups, from a very small amount of unlabelled training data.
In Section \ref{sec:low-latency-gcc-nmf}, we combine RT-GCC-NMF with an asymmetric STFT windowing approach to drastically reduce its inherent algorithmic latency.
In section \ref{subsec:enhancement-quality-experiments}, we show that we may reduce algorithmic latency from 80 ms to as low as 2 ms with no loss in speech enhancement performance, therefore achieving algorithmic latencies well within the tolerable range for hearing assistive devices.
In Section \ref{sec:experiments}, we study the effects of the RT-GCC-NMF system parameters and compare it with other approaches from the SiSEC challenge \cite{liutkus20172016}.
In Section \ref{sec:realtime-implementation}, we present an open source implementation of RT-GCC-NMF.
We evaluate the computational performance on a wide variety of hardware platforms and show that latencies as low as 6 ms are possible in practice.

\section{Offline GCC-NMF}\label{sec:offline-gcc-nmf}

\subsection{GCC: Generalized cross-correlation}

GCC is a robust approach to sound source localization in the presence of noise, interference, and reverberation \cite{knapp1976generalized, XAngueraThesis2006}.
The GCC function extends the frequency domain cross-correlation definition with an arbitrary frequency-weighting function $\psi_{ft}$, providing control over the relative importance of the signal's constituent frequencies when computing the cross-correlation:

\begin{equation}
\mathrm{G}_{\tau t}=\operatorname{Re} \sum_{f} \psi_{ft} \mathrm{V}_{Lft} \mathrm{V}_{Rft}^{*} e^{j2\pi f\tau}
\end{equation}
where $\mathrm{V}_{Lft}$ and $\mathrm{V}_{Rft}$ are the left and right complex-valued time-frequency transforms computed with the STFT, $^*$ is complex conjugation, and $f$, $t$, and $\tau$ index frequency, time, and TDOA respectively.\\
\indent Many of the most robust localization methods are based on the GCC phase transform (GCC-PHAT) \cite{carter1987coherence, omologo1994acoustic, brandstein1997robust}, in which frequencies are weighted equally by defining $\psi^{\mathrm{PHAT}}_{ft}$ as the reciprocal of the product of the magnitude spectrograms, i.e. $\psi^{\mathrm{PHAT}}_{ft}=\left(\left|\mathrm{V}_{Lft}\right|\left|\mathrm{V}_{Rft}\right|\right)^{-1}$, such that:
\begin{align}
\mathrm{G}^{\mathrm{PHAT}}_{\tau t} &= \operatorname{Re} \sum_{f} \frac{\mathrm{V}_{Lft}\mathrm{V}_{Rft}^{*}}{\left|\mathrm{V}_{Lft}\right|\left|\mathrm{V}_{Rft}\right|} e^{j2\pi f\tau}\\
&=\operatorname{Re} \sum_{f} e^{j\left( \angle \mathrm{V}_{Lft} - \angle \mathrm{V}_{Rft} \right)} e^{j2\pi f\tau}
\label{eq:gcc-phat}
\end{align}
The resulting GCC-PHAT \emph{angular spectrogram} can then be pooled over time, with the TDOA of the highest peaks corresponding to the source location estimates; see Figure \ref{fig:gcc-nmfs}b) for an example.

\subsection{NMF: Non-negative matrix factorization}\label{subsec:nmf-nonnegative-matrix-factorization}
\begin{figure}[t!]
\centering
\includegraphics{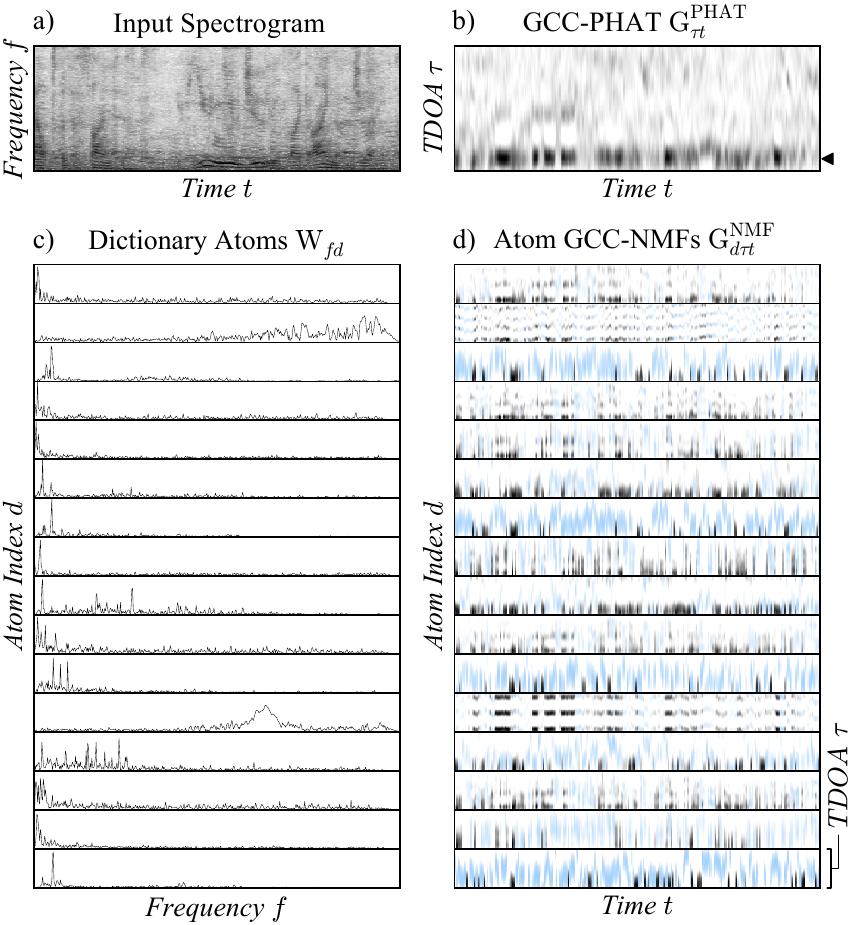}
\caption{Elements of the GCC-NMF speech enhancement algorithm for a 2-second mixture of speech and noise.
a) The input magnitude spectrogram, averaged over the left and right channels for presentation.
b) The GCC-PHAT angular spectrogram, with resulting target TDOA estimate indicated with a triangle marker.
c) Small subset of the 1024 NMF dictionary atoms $\mathrm{W}_{fd}$, with corresponding GCC-NMF angular spectrograms $\mathrm{G}^{\mathrm{NMF}}_{d\tau}$ shown in d).
Each row depicts the GCC-NMF angular spectrogram (TDOA $\tau$ vs. time $t$) for a given NMF dictionary atom.
GCC-NMF time frames for which an atom is associated with the target (see Section \ref{subsec:offline-gcc-nmf}) are colored in black, otherwise it is colored in light blue.
The angular spectrogram in b) is rectified for clarity with $\mathrm{max}(0,x)$, while each angular spectrogram in d) is each rectified using its median value with $\mathrm{max}(\mathrm{median}(x),x).$}
\label{fig:gcc-nmfs}
\end{figure}

When applying NMF to audio signals, input typically consists of a magnitude spectrogram $\left|\mathrm{V}_{ft}\right|$, with $f$ and $t$ indexing frequency and time as above.
NMF decomposes the spectrogram into two non-negative matrices: a dictionary $\mathrm{W}_{fd}$ whose columns comprise atomic spectra indexed by $d$ and set of corresponding activation coefficients $\mathrm{H}_{dt}$ such that $|\mathrm{V}|\approx\mathrm{W}\mathrm{H}$; see Figure \ref{fig:gcc-nmfs}c) for example NMF dictionary atoms.
Each column of the input spectrogram $|\mathrm{V}_{ft}|$, i.e. each frame $t$, is thus approximated as a linear combination of the NMF dictionary atoms with the activation coefficients from the corresponding column of $\mathrm{H}$.
For the stereo spectrograms we study here, we concatenate the left and right input spectrograms along the time axis, $\mathrm{V}=\left[ \mathrm{V}_{L} | \mathrm{V}_{R} \right]$, i.e. for left and right spectrograms each with size F x T, the concatenated matrix has size F x 2T.
In this way, the resulting NMF dictionary atoms capture only spectral information as before, with differences between the left and right channels captured in the corresponding activation coefficient matrices, $\mathrm{H}=\left[ \mathrm{H}_{L}| \mathrm{H}_{R}\right]$.\\
\indent In traditional NMF, dictionary learning and activation coefficient inference are performed concurrently by initializing the dictionary and activation coefficient matrices randomly, and updating them iteratively according to multiplicative update rules.
The update rules converge to a local minimum of the beta divergence reconstruction cost function, a special case of which is the generalized Kullback-Leibler (KL) divergence \cite{fevotte2011algorithms} defined as,
\begin{equation}
D\left(\left|\mathrm{V}\right|,\mathrm{\Lambda}\right) = \mathrm{|V|}\left(\log\mathrm{|V|}-\log\mathrm{\Lambda}\right)+\left(\mathrm{\Lambda}-\mathrm{|V|}\right)
\end{equation}
where $\Lambda=\mathrm{W}\mathrm{H}$ is the reconstructed input matrix $\mathrm{V}$.
The update rules for the KL divergence cost function are then \cite{lee2001algorithms},
\begin{equation}
\mathrm{H}\leftarrow\mathrm{H}\odot\mathrm{\frac{W^{\top}\frac{\left|\mathrm{V}\right|}{\mathrm{\Lambda}}}{\mathrm{W}\cdot 1}}\label{eq:nmf-update1}
\end{equation}
\begin{equation}
\mathrm{W}\leftarrow\mathrm{W}\odot\frac{\frac{\left|\mathrm{V}\right|}{\mathrm{\Lambda}}\mathrm{H}^{\top}}{1 \cdot \mathrm{H}^{\top}}\label{eq:nmf-update2}
\end{equation}
where the matrix exponentials, divisions, and Hadamard product $\odot$ are computed element-wise, and ${\bm1}$ is the all-ones matrix.
The NMF dictionary atoms are typically normalized after each update, and their activation coefficients scaled accordingly.\\
\indent Since all time frames (all columns of $\mathrm{V}_{ft}$) are required prior to optimization, standard NMF is an offline approach.
For RT-GCC-NMF, as described in Section \ref{subsec:coefficient-inference}, we will instead pre-learn the NMF dictionary and infer its activation coefficients online on a frame-by-frame basis by initializing the activation coefficient vector randomly and iteratively performing \eqref{eq:nmf-update1} while keeping the dictionary fixed.

\subsection{GCC-NMF}\label{subsec:offline-gcc-nmf}
Given the arbitrary frequency-weighting function $\psi_{ft}$ in the definition of GCC, and the fact that individual NMF dictionary atoms are themselves non-negative functions of frequency, we may construct a set of atom-specific GCC frequency weighting functions, 
\begin{equation}
\psi_{dft}^{\mathrm{NMF}}=\frac{1}{\left| \mathrm{V}_{Lft} \right| \left| \mathrm{V}_{Rft} \right|}\frac{\mathrm{W}_{fd}}{\sum_{f}\mathrm{W}_{fd}}
\end{equation}
such that for a given atom $d$, frequencies are weighted according to their relative magnitude in the atom.
The resulting atom-specific GCC-NMF angular spectrograms are then defined as follows, with examples shown in Figure \ref{fig:gcc-nmfs}d),
\begin{equation}
\mathrm{G}_{d\tau t}^{\mathrm{NMF}}= \operatorname{Re} \sum_{f}\psi_{dft}^{\mathrm{NMF}} \mathrm{V}_{Lft}\mathrm{V}_{Rft}^{*} e^{j2\pi f\tau}
\end{equation}
These GCC-NMF angular spectrograms then allow us to estimate the TDOA of each atom $d$ at each time $t$, defined as the $\tau$ for which GCC-NMF reaches its maximum value, i.e. $\text{argmax}_{\tau}\mathrm{G}_{d\tau t}^{\mathrm{NMF}}$.
We then associate individual atoms with the target if their estimated TDOA lies within a window of size $\epsilon$ around the target TDOA $\hat{\tau}_t$ as estimated by GCC-PHAT, otherwise they are associated with interference.
This procedure defines a binary activation coefficient mask,
\begin{equation}
\bar{\mathrm{M}}_{dt}=\begin{cases}
1 & \text{if }|\hat{\tau}_t-\text{argmax}_{\tau}\mathrm{G}_{d\tau t}^{\mathrm{NMF}}| < \epsilon / 2\\
0 & \text{otherwise}
\end{cases}
\label{eq:mask-generation}
\end{equation}
where the effect of the window width $\epsilon$ will be studied in Section \ref{sec:experiments} in the context of the soft-masking alternative presented below.
Multiplying the mask $\mathrm{M}_{dt}$ with the activation coefficients $\mathrm{H}_{dt}$ element-wise and reconstructing as usual then yields a primary estimate of the target magnitude spectrogram,
\begin{equation}
|\bar{\mathrm{X}}_{cft}|=\sum_{d}\mathrm{W}_{fd}\mathrm{H}_{cdt}\bar{\mathrm{M}}_{dt}
\label{eq:target-magnitude-spectrogram}
\end{equation}
We note that this mask eliminates atoms attributed to the interference, thus isolating the target speech from the mixture.
As is typical in NMF-based separation, the complex target spectrogram is then estimated by applying a time-varying Wiener-like filter to the input signal.
This filter is constructed in the frequency domain as the ratio between the target and mixture estimate spectrograms $\mathrm{\Lambda}_{cft}=\sum_d W_{fd}H_{cdt}$, i.e. the reconstructed estimate of the magnitude input spectrogram $|\mathrm{V}_{cft}|$.
The filter is then multiplied with the complex input spectrogram $\mathrm{V}_{cft}$,
\begin{equation}
\hat{\mathrm{X}}_{cft}=\frac{|\bar{\mathrm{X}}_{cft}|}{\mathrm{\Lambda}_{cft}}\mathrm{V}_{cft}
\label{eq:wiener-filtering}
\end{equation}
where $\hat{\mathrm{X}}_{cft}$ is the complex target spectrogram estimate and $c$ is the channel index.
The complex target spectrogram estimate is then transformed to the time domain with the inverse STFT as described in Section \ref{sec:stft-and-latency}.

\subsection{Soft masking GCC-NMF}\label{subsec:soft-gcc-nmf-masking}
Soft-mask alternatives to binary masking in the \emph{time-frequency} domain is a common technique to improve speech enhancement performance \cite{radfar2007single, reddy2007soft}.
In this section, we propose a soft-mask alternative to the binary \emph{activation coefficient mask} we reviewed above.
This soft NMF activation coefficient masking function is defined as,
\begin{equation}
\widetilde{\mathrm{M}}_{dt}=\left(1-\eta \right)\exp\left(-\left(\frac{\left|\hat{\tau}_t - \textrm{argmax}_\tau{\textrm{G}_{d\tau t}^{\textrm{NMF}}}\right|}{\alpha}\right)^{\beta}\right) + \eta
\label{eq:soft-mask-generation}
\end{equation}
where $d$ is the atom index, $\hat{\tau}_t$ is the target TDOA, $\alpha$ controls the window width, $\beta$ controls the window shape, and $\eta$ defines the window floor, i.e. its minimum value.
This soft mask allows atoms to be attenuated in a continuous fashion based on the distance between their estimated TDOA and the target TDOA.
In Section \ref{subsec:experiments-window-parameters}, we will study the effect of the masking function parameters of the binary and soft activation coefficient masks on objective speech enhancement quality and speech intelligibility measures.
We will show that the parameters may be used to control the trade-off between interference suppression and target fidelity, as well as the trade-off between speech quality and speech intelligibility.
For the other experiments presented in Section \ref{sec:experiments}, we set the masking parameters to $\alpha$ = 3/16, $\eta$ = 0, and $\beta=\infty$, for which the soft masking function reduces to the binary masking function.
Since the resulting mask is applied independently for each time frame, the parameters may be modified in real-time based on the user's needs.

\section{RT-GCC-NMF}\label{sec:online-gcc-nmf}

As the GCC-NMF masking functions defined in the previous section are constructed independently for each frame, GCC-NMF has \emph{potential} to be performed online in a frame-by-frame fashion.
However, dictionary learning, activation coefficient inference, and target localization are performed using the entire mixture signal, thus precluding online use.
We proceed to address each of these elements in this section, as we develop the real-time RT-GCC-NMF.

\begin{figure}[t!]
\centering\includegraphics[width=\linewidth]{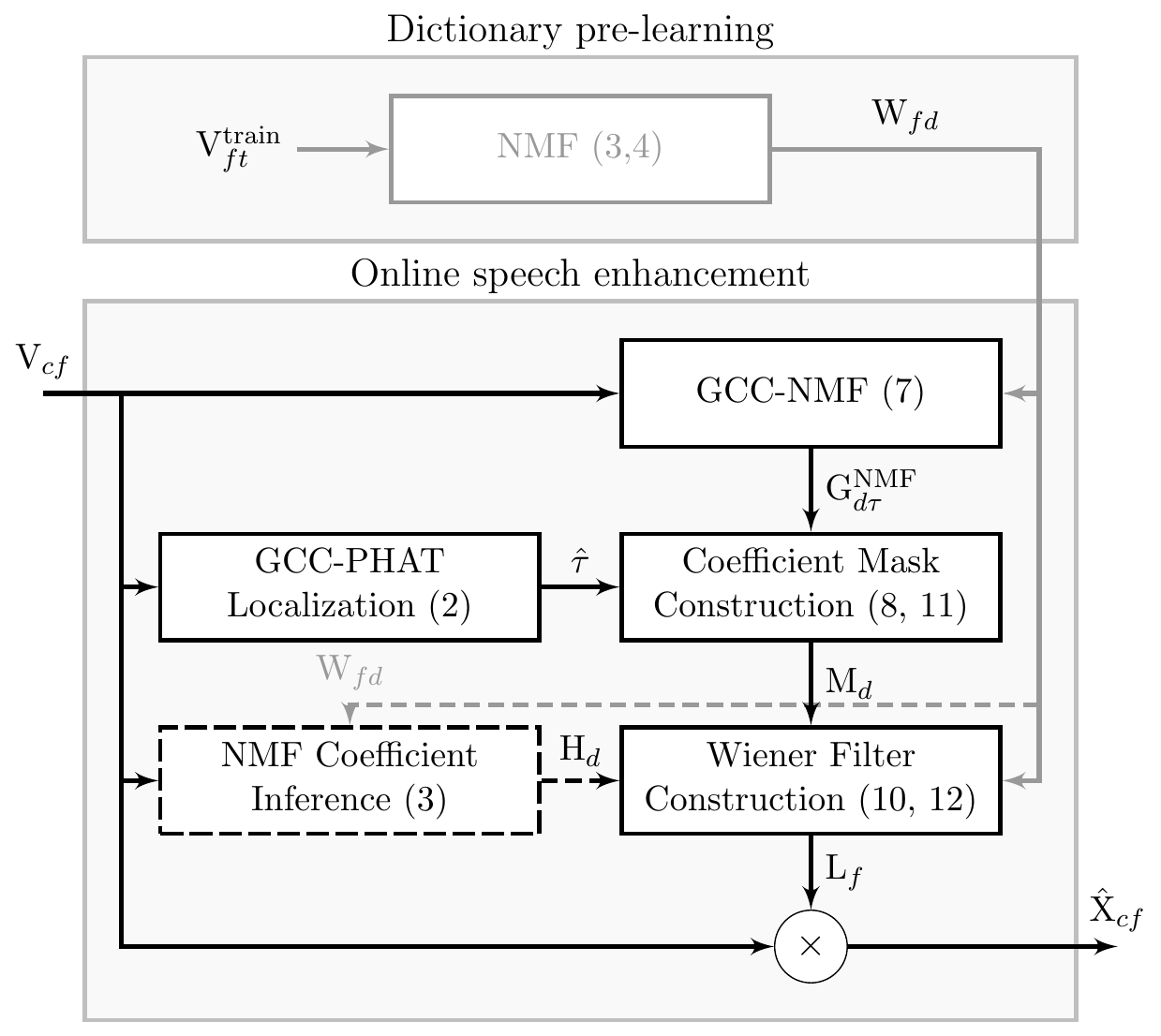}
\caption{Block diagram of RT-GCC-NMF consisting of offline dictionary pre-learning and online speech enhancement.
Online, offline, and optional components are drawn with black, gray, and dotted lines respectively, with relevant equations for each block listed in parentheses.}
\label{fig:BlockDiagram}
\end{figure}

\subsection{Dictionary pre-learning}\label{sec:dictionary-pre-learning}
A typical approach for \emph{supervised} speech enhancement with NMF is to pre-learn a pair of NMF dictionaries: one using isolated speech and one using isolated noise.
For a given test signal, the activation coefficients of both dictionaries are then inferred while keeping the dictionaries fixed \cite{weninger2014discriminative, vincent2014blind}.
We adapt this approach to the \emph{unsupervised} setting here by pre-learning a \emph{single} NMF dictionary from a dataset containing \emph{both} isolated speech and noise signals.
As described in Section \ref{sec:experimental-setup}, individual STFT frames are chosen at random from the isolated speech and isolated noise signals from the CHiME dataset.
These frames are concatenated along the time axis to construct an NMF input matrix $\mathrm{V}$ that is used as input to a standard NMF decomposition, i.e. using equations {\eqref{eq:nmf-update1}} and {\eqref{eq:nmf-update2}}.
We keep the resulting NMF dictionary $\mathrm{W}$, comprising elements of both speech and noise, and the resulting activation coefficients $\mathrm{H}$ are discarded.\\
\indent Contrary to the supervised approach, this approach remains purely unsupervised as a single dictionary is learned for both speech and noise using no prior knowledge.
As the single pre-learned NMF dictionary contains features of both speech and noise signals, individual NMF dictionary atoms are then associated with the target speaker or interference at each point in time according to \eqref{eq:mask-generation} or \eqref{eq:soft-mask-generation}.
This approach allows individual NMF dictionary atoms to encode either speech or noise at different points in time, thus overcoming the limitation in the supervised case where a single dictionary atom may only encode a single source.
In Section \ref{subsec:dictionary-pretraining}, we show that we may achieve comparable performance to offline GCC-NMF by pre-learning the NMF dictionary using one dataset and testing on a completely different dataset with different speakers, acoustic environments, and recording setups.
The dictionary pre-learning approach is therefore able to generalize across these conditions, avoiding the well-known mismatch problem with NMF-based speech enhancement when the training and testing data originate from different datasets \cite{mohammadiha2013supervised}.

\subsection{Activation coefficient inference}\label{subsec:coefficient-inference}
The activation coefficients of the pre-learned dictionary can be inferred for the input mixture on a frame-by-frame basis by initializing the activation coefficient vector randomly, and updating it iteratively according to \eqref{eq:nmf-update1}.
We note that since the estimated target is $\mathrm{W}\left(\mathrm{H}\odot\mathrm{M}\right)$ and the estimated interference is $\mathrm{W}\left(\mathrm{H}\odot\left(\mathrm{1-M}\right)\right)$, the estimated mixture is $\mathrm{W}\mathrm{H}$ (the sum of target plus interference).
Inference of the mixture coefficients $\mathrm{H}$ is therefore performed independent of the coefficient mask $\mathrm{M}$ estimation.
The coefficient mask then attenuates atoms that are attributed to noise based on their TDOA estimates.
We will see in Section \ref{sec:number-of-updates} that better performance can in fact be achieved by forgoing activation coefficient inference altogether.
In this case, we may replace the activation coefficients $\mathrm{H}_{dt}$ with all-ones, thus simplifying the Wiener-like filtering process defined in \eqref{eq:wiener-filtering} as follows,
\begin{align*}
\hat{\mathrm{X}}_{cft}&=\frac{|\bar{\mathrm{X}}_{cft}|}{\mathrm{\Lambda}_{cft}}\mathrm{V}_{cft}\\
&=\frac{\sum_d\mathrm{W}_{fd}\mathrm{H}_{cdt}\mathrm{M}_{dt}}{\sum_d\mathrm{W}_{fd}\mathrm{H}_{cdt}}\mathrm{V}_{cft}\\[0.5em]
       &=\frac{\sum_d\mathrm{W}_{fd}\mathrm{M}_{dt}}{\sum_d\mathrm{W}_{fd}}\mathrm{V}_{cft} \numberthis \label{eq:wiener-filtering-no-inference}
\end{align*}
making use of the definition of $|\bar{\mathrm{X}}_{cft}|$ from Eq. \eqref{eq:target-magnitude-spectrogram} and $\Lambda_{cft}$ as defined in text thereafter.
An interesting consequence of this simplification is that the resulting filter no longer relies on the input magnitudes $\left|\mathrm{V}_{cft}\right|$ that were used to infer the NMF activation coefficients $\mathrm{H}_{cdt}$.
Depending only on the pre-learned dictionary and phase differences between the left and right channels, this simplification results in a purely \emph{phase-based} variant of RT-GCC-NMF.
As well, since a single mask is used for both channels, binaural cues remain unaffected by the filtering process.

\subsection{Online localization}\label{subsec:online-localization}
With offline GCC-NMF, target localization was performed using a max-pooled GCC-PHAT technique \cite{blandin2012multi} where the target TDOA is that at which the global maximum occurred in the GCC-PHAT angular spectrogram defined in \eqref{eq:gcc-phat}, i.e. $\text{argmax}_{\tau}\text{max}_t\mathrm{G}_{\tau t}^{\mathrm{PHAT}}$.
In the online setting, however, we only have access to present and past information and the target localization method must consequently be adapted.
We explore two online localization approaches here, for both static and moving speakers.
In Section \ref{sec:experiments}, we consider the static speaker scenario.
In this case, we take the current and all previous angular spectrogram frames into account in taking the argmax, i.e. $\text{argmax}_{\tau}\text{max}_{t'}\mathrm{G}_{\tau t'}^{\mathrm{PHAT}}$ for $t'\leq t$.
In Section \ref{sec:realtime-implementation}, we extend this approach to the moving speaker case with a real-time open source demonstration.
In this case, we use a sliding window approach where the argmax is taken over the recent history of the angular spectrogram, i.e. $\text{argmax}_{\tau}\text{max}_{t'}\mathrm{G}_{\tau t'}^{\mathrm{PHAT}}$ for $t-L\leq t'\leq t$,
where $L$ is the sliding window size.
The effect of the window size may be explored interactively in real-time, where smaller window sizes track faster changes in source position but may switch to background noise during short pauses in the speech, while larger window sizes result in a more stable tracking for more slowly moving speakers.

\section{Low latency RT-GCC-NMF}\label{sec:low-latency-gcc-nmf}
Speech enhancement algorithms built around the STFT incur an inherent algorithmic latency, independent of processing speed, equal to the window size plus the hop size (frame advance).
Given the trade-off between spectral resolution and window size, algorithms including RT-GCC-NMF that rely on high spectral resolution often have latencies greater than 64 ms.
However, such high latencies are not tolerable for a number of real-world applications of speech enhancement including assistive listening devices.
In this section, we combine the asymmetric STFT windowing approach proposed by Mauler and Martin \cite{mauler2007low} with the RT-GCC-NMF speech enhancement system, thus simultaneously providing high spectral resolution and latencies well below the 10 ms target for hearing aids.

\subsection{STFT and latency}\label{sec:stft-and-latency}
The STFT processes sound in frames, i.e. short overlapping segments of time, where each frame is multiplied by an \emph{analysis window} prior to computing its Fourier transform.
Resynthesis is achieved by taking the inverse Fourier transform of the transformed frame, multiplying the resulting samples by a \emph{synthesis window}, and combining neighbouring frames via the overlap-add (OLA) method \cite{smith2011spectral}.
Perfect reconstruction can be achieved if the transform has the constant overlap-add (COLA) property, i.e. if the overlapped sum of the element-wise product of the analysis and synthesis windows is constant over time.
A commonly-used window for analysis and synthesis is the pointwise square root of the periodic Hann window \cite{oppenheim1999discrete}, where the periodic Hann function is defined for frame size $N$ as,
\begin{equation}
\mathrm{H}_{N}[n]=\begin{cases}
\frac{1}{2}\left(1-\cos\left(2\pi\frac{n}{N}\right)\right) & 0 \leq n < N\\
0 & \text{otherwise}
\end{cases}
\end{equation}
\indent The above process of overlapped signal windowing with OLA resynthesis induces a latency $\mathrm{L}_\mathrm{OLA}$ equal to the window size $N$.
To run in real-time, all processing including the Fourier transform and its inverse, should occur within a single frame advance $R$, resulting in a total system latency of $N+R$.
Running RT-GCC-NMF on input signals sampled at 16 kHz, with a window size of 1024 samples and 256 sample frame advance, for example, results in a total system latency of 80 ms.\\
\indent A first approach to reduce the RT-GCC-NMF system latency is to simply reduce the window size $N$.
This comes at the expense of decreasing the spectral resolution, however, and as we will show in Section \ref{subsec:enhancement-quality-experiments}, objective speech enhancement quality and intelligibility measures decrease significantly for small window sizes with this approach.
We therefore propose an alternative approach to latency reduction based on an asymmetric STFT windowing method.
\begin{figure}[t!]
\centering
\includegraphics{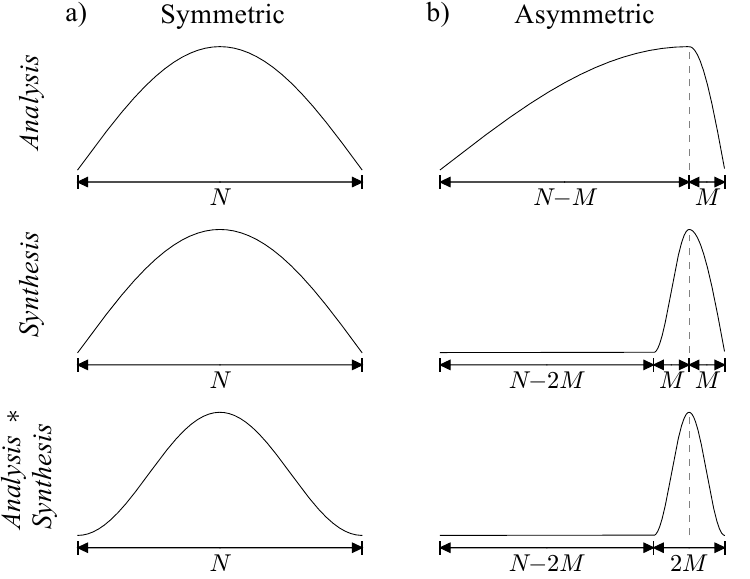}
\caption{Comparison of the symmetric and asymmetric STFT window functions for frame size $N$.
a) Traditional symmetric square root Hann analysis and synthesis window functions and their element-wise product Hann window.
b) Asymmetric window functions, where the analysis window has duration $N$ and is weighted towards the right, while the synthesis window has duration $2M{<}N$, and shares its right edge with the underlying frame.
The resulting pointwise product of the analysis and synthesis windows is a Hann window of size $2M$ that also shares its right edge with the underlying frame.}
\label{fig:windows-figure}
\end{figure}
\subsection{Asymmetric STFT windowing}\label{sec:asymmetric-stft-windowing}
Departing from the tradition of symmetric analysis and synthesis windows that have the same duration, asymmetric windowing allows us to simultaneously achieve high spectral resolution and low latency by combining long analysis windows with short synthesis windows.
The asymmetric windows we use in this work have been adapted from \cite{mauler2007low}, though other asymmetric windowing approaches can be found in the literature \cite{allamanche1999mpeg, schnell2008mpeg, su2016minimum, andersen2016adaptive}.\\
\indent For a given frame size $N$, the asymmetric analysis and synthesis windows are designed such that their product is a Hann window of size $2M<N$.
This Hann window shares its right edge with the underlying frame, and can be made to be much shorter than the frame itself by choosing $2M\ll N$, as depicted in Figure \ref{fig:windows-figure}.
The analysis window $h_{A}$ and the synthesis window $h_{S}$ are defined mathematically as,
\begin{equation}
h_{A}[n]=\begin{cases}
\sqrt{\mathrm{H}_{2\left(N{-}M\right)}}[n] & 0 \leq n < N{-}M\\
\sqrt{\mathrm{H}_{2M}}[n{-}\left(N{-}2M\right)] & N{-}M \leq n < N\\
0 & \text{otherwise}
\end{cases}
\end{equation}
\begin{equation}
h_{S}[n]=\begin{cases}
\frac{ \mathrm{H}_{2M}[n{-}(N{-}2M)]}{ \sqrt{\mathrm{H}_{2\left(N{-}M\right)}}[n] } & N{-}2M \leq n < N{-}M\\
\sqrt{\mathrm{H}_{2M}}[n{-}(N{-}2M)] & N{-}M \leq n < N\\
0 & \text{otherwise}
\end{cases}
\end{equation}
\noindent These window functions are constructed in two parts with respect to the center of the analysis-synthesis product Hann window, i.e. $n=N{-}M$.
To the right of $N{-}M$, both analysis and synthesis windows consist of the right half of a square root Hann window of size $2M$.
To the left, the analysis window consists of the left half of a Hann window of size $N{-}M$, while the synthesis window is defined as the ratio of the analysis window and the product Hann window, limited to the range $N{-}2M \leq n < N{-}M$.\\
\indent In Section \ref{sec:experiments-low-latency}, we will show that this asymmetric windowing strategy allows us to drastically reduce the algorithmic latency of RT-GCC-NMF without affecting objective speech enhancement quality and intelligibility metrics.
For example, running RT-GCC-NMF on 16kHz input signal with an analysis window of 1024 samples and a synthesis window of 32 samples with a 16 sample frame advance, we may reduce the total system latency to 3 ms.
We note that retaining the relative synthesis window overlap while decreasing the synthesis window size results in an increase in the number of windows required for the overlap-add process, and thus increases the computational load.
We present an empirical analysis of the computational requirements for varying algorithmic latencies in Section \ref{subsec:latency-processing-time}.
Finally, we note that the asymmetric windowing approach increases the start-up latency of the STFT, though this may be mitigated by pre-padding the signal with zeros.
\begin{figure}[t]
	\centering
	\includegraphics{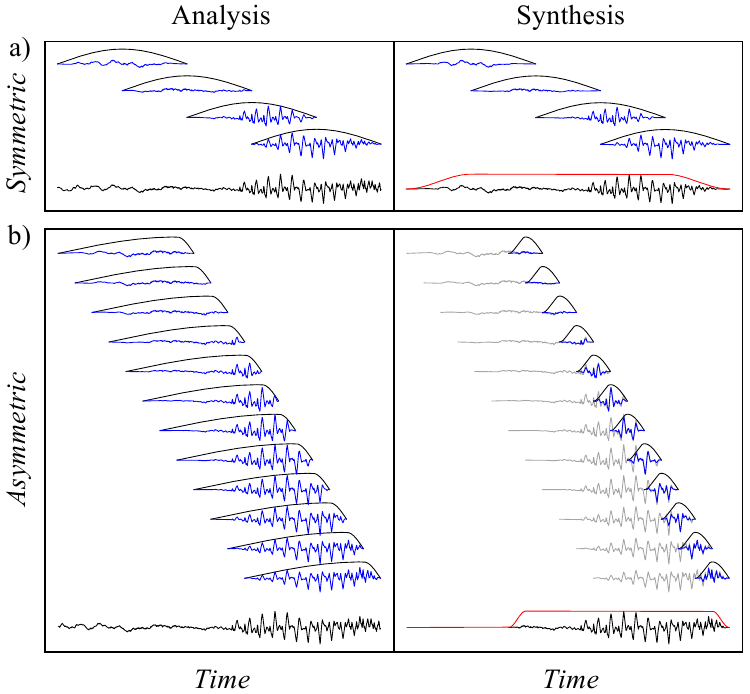}
	\caption{Comparison of overlap-add procedures for the a) symmetric and b) asymmetric STFT windowing methods.
	Note that while the synthesis windows are mostly zero in the asymmetric case, perfect reconstruction is still achieved when the frame advance is sufficiently small to allow for constant overlap-add of the window functions (shown in red).}
	\label{fig:ola-analysis-synthesis}
\end{figure}
\section{Experiments}\label{sec:experiments}

\subsection{Experimental setup}\label{sec:experimental-setup}

In this section, we evaluate the RT-GCC-NMF algorithm on the SiSEC 2016 speech in noise \emph {dev} dataset, consisting of two-channel mixtures of speech and real-world background noise, with microphones separated by 8.6 cm \cite{liutkus20172016}.
Unsupervised dictionary pre-learning is performed on a small subset of the CHiME 2016 development set \cite{vincent2016analysis}, with randomly selected frames equally divided between isolated speech and background noise signals of a single microphone.
The sample rate for both SiSEC and CHiME is 16 kHz, and we use an STFT with 1024-sample windows (64 ms), a 256-sample hop size (16 ms), and a square root Hann analysis and synthesis window functions for the symmetric windowing case.
Default RT-GCC-NMF parameters are set to dictionary size = 1024, number of NMF dictionary pre-learning updates = 100, number of NMF activation coefficient inference updates at runtime = 100, number of TDOA samples = 128, and target TDOA window size 3/64 of the total range, i.e. 6 TDOA samples.\\
\indent Speech enhancement quality is quantified using the Perceptual Evaluation methods for Audio Source Separation (PEASS) toolkit \cite{emiya2011subjective}, and the BSS Eval ``toolbox for performance measurement in (blind) source separation" \cite{vincent2006performance}.
PEASS is a perceptually-motivated method that better correlates with subjective assessments than the traditional SNR-based metrics provided by BSS Eval \cite{emiya2011subjective}.
These open source toolkits both provide measures of overall enhancement quality, target fidelity, interference suppression, and lack of perceptual artifacts, where higher scores are better in all cases.
The overall, target-related, interference-related, and artifact-related scores are named OPS, TPS, IPS, and APS respectively in the case of PEASS, and SDR, ISR, SIR, and SAR in the case of BSS Eval.
Speech intelligibility is quantified with the short-time objective intelligibility (STOI) \cite{taal2011algorithm} and the extended STOI (ESTOI)  \cite{jensen2016algorithm} measures, where ESTOI has been shown to better correlate with listening test scores than STOI \cite{van2017evaluation}.
\begin{figure}[t!]
\centering\includegraphics[]{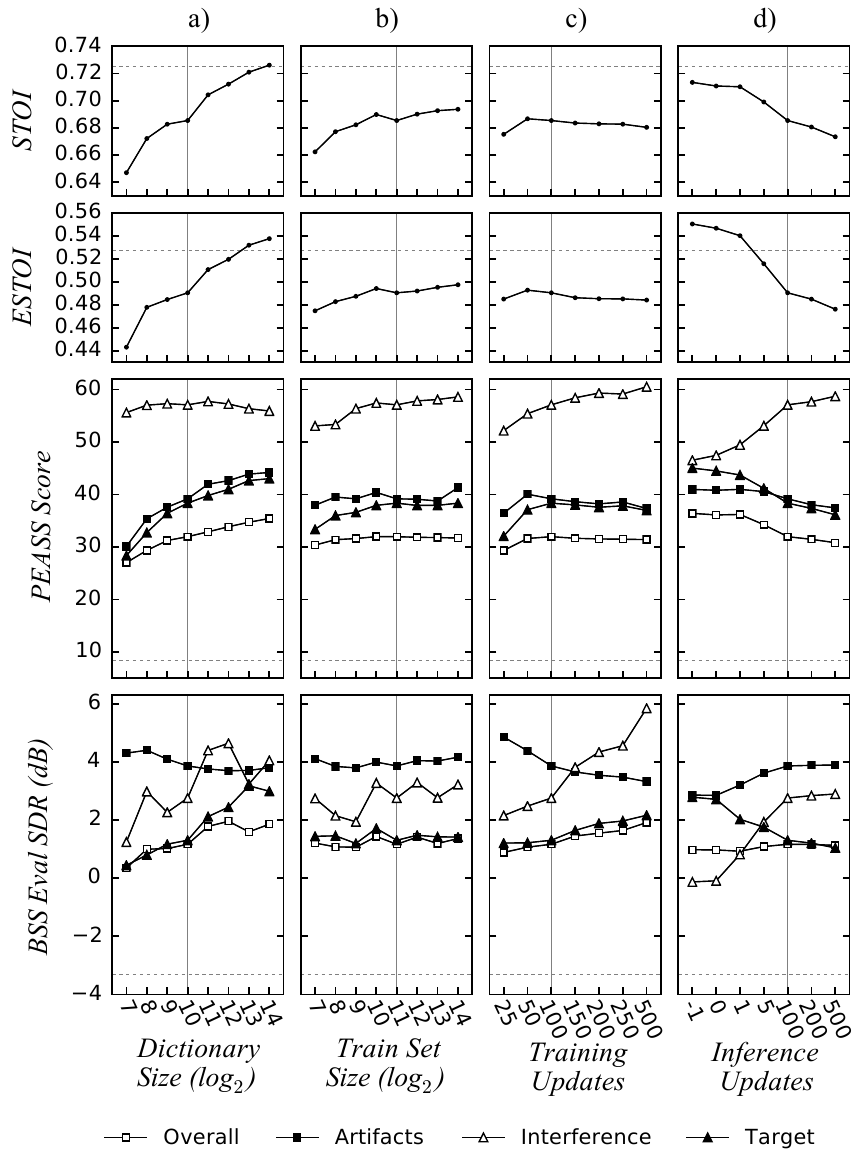}
\caption{Effect of various RT-GCC-NMF parameters on BSS Eval, PEASS, ESTOI, and STOI metrics averaged over the SiSEC speech in noise \emph{dev} dataset for a) NMF dictionary size and b) number of frames used for dictionary pre-learning, both varying from $2^7$ (128) to $2^{14}$ (16 384) exponentially; c) number of NMF pre-learning update iterations; d) number of NMF activation coefficient inference updates performed at runtime.
For PEASS and BSS Eval scores, the four scores presented correspond to overall quality, target fidelity, interference suppression, and lack of artifacts, i.e. for PEASS: OPS, TPS, IPS, and APS, and for BSS Eval: SDR, ISR, SIR, and SAR.
Higher scores are better in all cases.
Default parameter values are indicated with vertical gray lines and average scores for the unprocessed mixture signals are indicated with horizontal dashed lines.}
\label{fig:train-dictionary-size}
\end{figure}

\subsection{RT-GCC-NMF experiments}\label{subsec:experiments-online}
We begin by studying the effects on objective speech enhancement quality and speech intelligibility metrics for RT-GCC-NMF.
We first study the effects of the pre-learned dictionary size and the amount of data used for pre-learning, followed by the number of training and inference iterations, the RT-GCC-NMF target TDOA masking function parameters, and the input SNR.
These evaluations are performed with offline target TDOA estimation using the max-pooled GCC-PHAT approach.
We then compare the offline localization approach with the simple accumulated online localization method, and compare the results with other speech enhancement algorithms from the SiSEC challenge and the ideal binary mask oracle baseline.
\begin{figure}[t!]
\centering\includegraphics{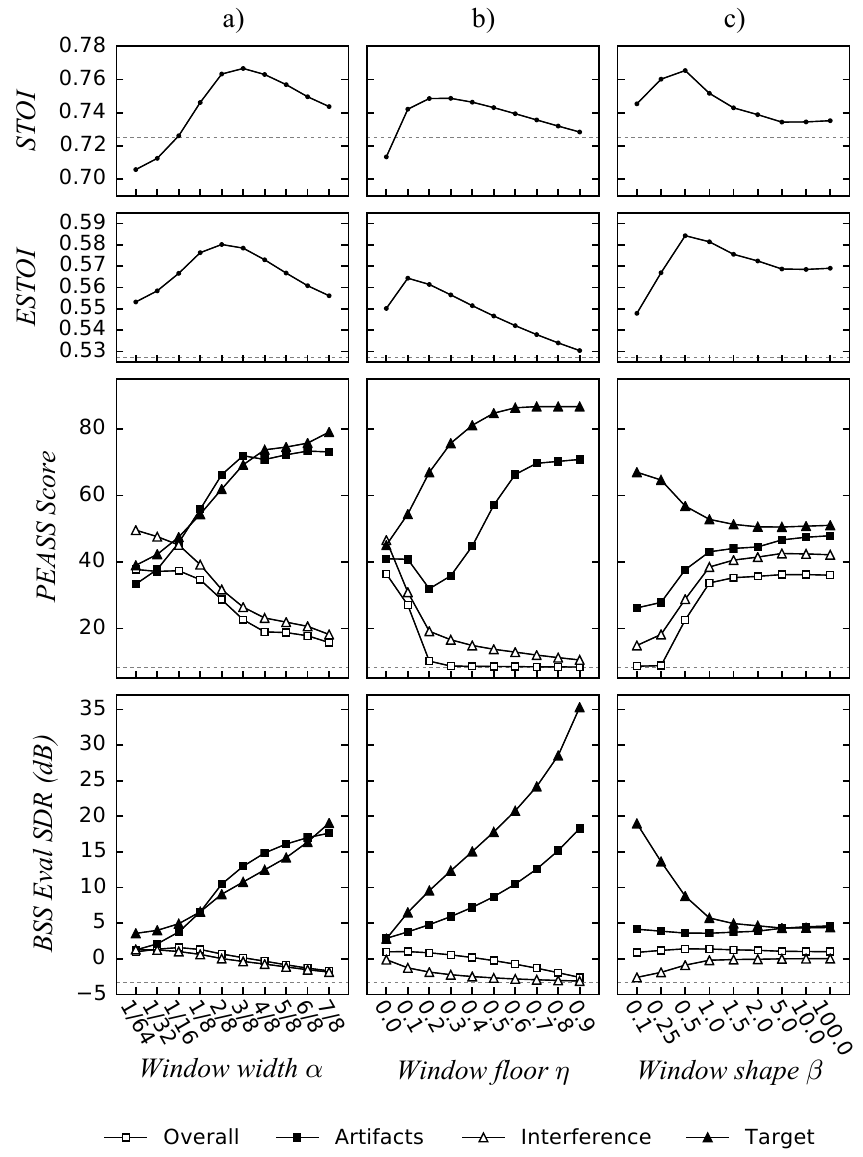}
\caption{Effect of RT-GCC-NMF masking function parameters on enhancement quality and objective intelligibility: a) TDOA window width, b) window floor, c) window shape as defined in \eqref{eq:soft-mask-generation}.
Scores are presented as in Figure \ref{fig:train-dictionary-size}, where dashed lines indicate the average scores for the unprocessed mixture signals.}
\label{fig:window-function-parameters}
\end{figure}
\begin{figure}[t!]
\centering\includegraphics{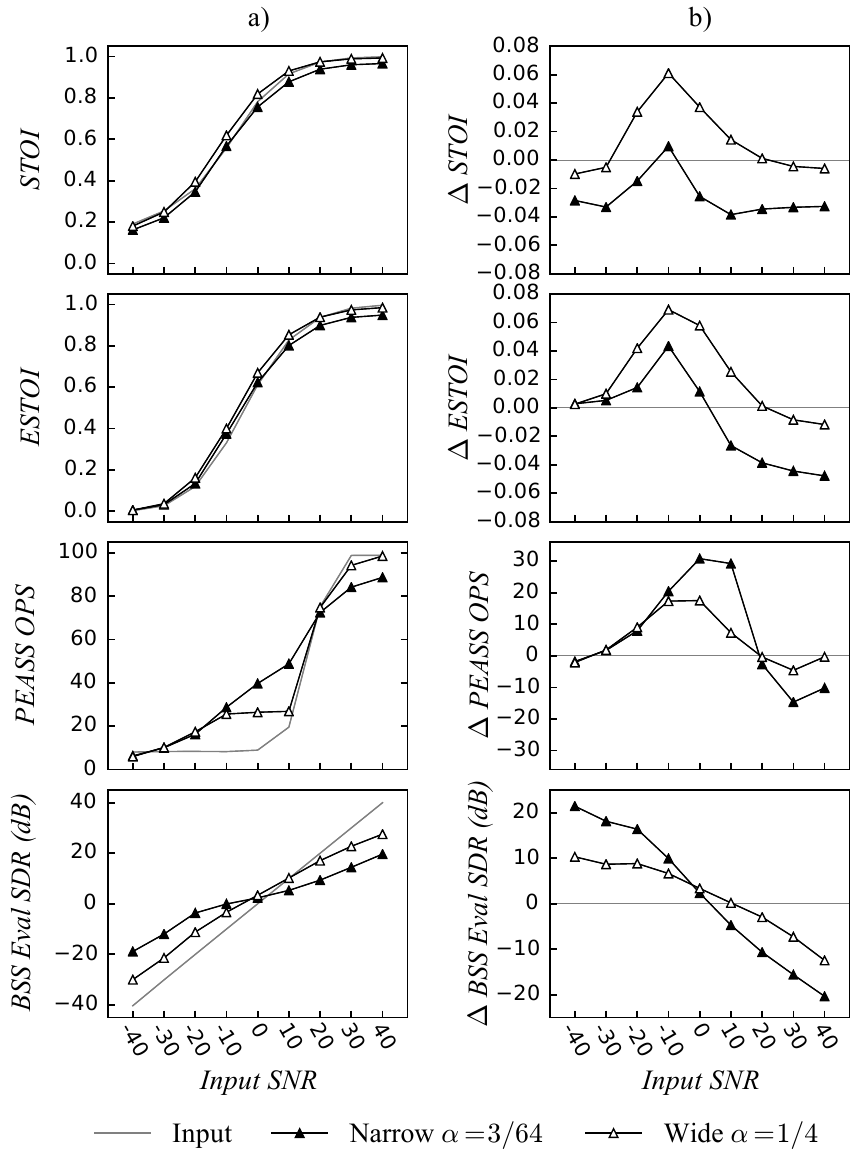}
\caption{Effect of input SNR on RT-GCC-NMF speech enhancement performance, for SNRs varying from -40 to 40 dB.
Results for both a narrow ($\alpha=3/64$) and wide ($\alpha=1/4$) TDOA window are shown.
Overall scores are presented for the PEASS and BSS Eval measures.
a) Absolute performance measures with triangle markers for the narrow window case, square markers, and the input score as a solid gray line.
Note the identity line in the case of BSS Eval SDR, where the input SNR and SDR evaluation metric are equivalent.
b) Performance measure improvements, with zero improvement emphasized with a gray line.}
\label{fig:effect-of-snr}
\end{figure}
\subsubsection{Dictionary pre-learning}\label{subsec:dictionary-pretraining}
Speech enhancement scores for varying dictionary size and number of frames used for dictionary pre-learning are shown in Figure \ref{fig:train-dictionary-size} a) and b) respectively, with default dictionary size of 1024, and default train set size of 2048 frames of 64 ms each, divided equally between speech and noise.
Overall PEASS and intelligibility scores converge quickly with increasing train set size, such that performance is near maximal for most measures with only 1024 training frames, while SNR-based scores exhibit more variability across train set sizes.
Contrary to many supervised approaches, therefore, the unsupervised dictionary pre-learning we present here requires a very small amount of training data.
For the NMF dictionary size, we note a monotonic increase in overall PEASS and intelligibility scores with diminishing returns.
While the SNR scores again exhibit more variability, they generally increase with dictionary size as well.
We observed a similar trend with offline GCC-NMF for increasing dictionary size, as well as similar overall PEASS and SNR scores on the same dataset \cite{wood2017blind}.
The dictionary pre-learning technique therefore generalizes to new speakers, noise and acoustic conditions, and recording setups.
In Section \ref{sec:Experiments-Online}, we study the capacity for generalization further by comparing sources used for dictionary learning in more detail.

\begin{table*}\centering \ra{1.2} 
\setlength{\tabcolsep}{2.5pt}\caption{Effects on speech enhancement scores of different elements of GCC-NMF. Mean scores $\pm$ standard deviation are taken over the SiSEC 2016 speech in noise \emph{dev} dataset with static speaker. For PEASS and BSS Eval, we present the OPS and SDR scores respectively.}
\begin{tabular}{ c @{\hskip 0pt} c @{\hskip 0pt} c @{\hskip 0pt} c @{\hskip 0pt} llll @{\hskip 0pt} c @{\hskip 0pt} llll } 
&&&& \multicolumn{4}{c}{1024 atoms} & \phantom{} & \multicolumn{4}{c}{16384 atoms} \\ 
\cmidrule{5-8} \cmidrule{10-13}
\multicolumn{1}{l}{Dictionary} & \multicolumn{1}{l}{Coefficients} & \multicolumn{1}{l}{Localization}& \phantom{abc} & \multicolumn{1}{c}{PEASS} & \multicolumn{1}{c}{BSS Eval} & \multicolumn{1}{c}{STOI} & \multicolumn{1}{c}{ESTOI}& \phantom{abc} & \multicolumn{1}{c}{PEASS} & \multicolumn{1}{c}{BSS Eval} & \multicolumn{1}{c}{STOI} & \multicolumn{1}{c}{ESTOI}\\ 
\midrule 
Mixture & Inferred & Offline & & 32.46$\pm$5.56 & 1.86$\pm$1.28 & 0.687$\pm$0.083 & 0.505$\pm$0.110 &  & 34.46$\pm$6.16 & 1.88$\pm$1.36 & 0.715$\pm$0.076 & 0.538$\pm$0.106\\Mixture & All-ones & Offline & & 35.82$\pm$5.15 & 1.60$\pm$1.06 & 0.698$\pm$0.069 & 0.537$\pm$0.089 &  & 37.96$\pm$4.52 & 1.41$\pm$1.18 & 0.715$\pm$0.067 & 0.561$\pm$0.081\\Pre-learned & Inferred & Offline & & 31.95$\pm$5.96 & 1.57$\pm$1.81 & 0.695$\pm$0.085 & 0.500$\pm$0.107 &  & 35.15$\pm$5.94 & 1.93$\pm$1.76 & 0.727$\pm$0.079 & 0.540$\pm$0.102\\Pre-learned & Inferred & Online & & 31.36$\pm$5.73 & 1.41$\pm$2.21 & 0.684$\pm$0.094 & 0.491$\pm$0.117 &  & 34.61$\pm$5.63 & 1.94$\pm$1.78 & 0.720$\pm$0.087 & 0.531$\pm$0.110\\Pre-learned & All-ones & Offline & & 36.69$\pm$4.82 & 1.16$\pm$1.19 & 0.716$\pm$0.060 & 0.551$\pm$0.082 &  & 37.58$\pm$3.74 & 1.52$\pm$0.98 & 0.720$\pm$0.059 & 0.563$\pm$0.080\\Pre-learned & All-ones & Online & & 35.80$\pm$4.87 & 1.05$\pm$1.30 & 0.704$\pm$0.058 & 0.541$\pm$0.084 &  & 35.87$\pm$3.57 & 1.40$\pm$1.06 & 0.711$\pm$0.060 & 0.556$\pm$0.081\\
 \bottomrule \end{tabular} \label{tab:gccNMFResultsTable} \end{table*}

\subsubsection{NMF training and inference updates}\label{sec:number-of-updates}

The effect of the number of NMF dictionary pre-learning updates on enhancement performance is presented in Figure \ref{fig:train-dictionary-size} c).
As was the case for offline GCC-NMF, increasing the number of training iterations results in increased interference suppression, with PEASS target fidelity and lack of artifact scores decreasing for higher number of iterations, while the overall PEASS and intelligibility scores remain stable for values greater than 50 updates.
The choice of the number of training iterations therefore offers offline control of the trade-off between target fidelity and interference suppression.
One could learn a set of dictionaries spanning a range of training iterations, and subsequently control the trade-off online by selecting the desired dictionary on a frame-by-frame basis.\\
\indent In Figure {\ref{fig:train-dictionary-size}} d), we present the effect of the number of online NMF inference iterations, i.e. the number of times equation {\eqref{eq:nmf-update1}} is called for a given input frame, after random initialization of the NMF activation coefficients $\mathrm{H}$.
We note similar convergence effects to the number of training iterations for large numbers of updates.
For small numbers of updates, however, we note an opposite effect for both overall PEASS and intelligibility scores, as they both increase with decreasing iterations.
The best overall PEASS and intelligibility scores is in fact achieved when no inference is performed, i.e. when random activation coefficients are used.
As mentioned in Section \ref{subsec:coefficient-inference}, we can thus forego the activation coefficient inference stage completely, and perform the Wiener-like filtering using only the pre-learned dictionary and input phase differences as in equation \eqref{eq:wiener-filtering-no-inference}.
These results are presented as well, i.e. using all-ones activation coefficients indicated with inference updates = -1, showing slightly better results than for random activation coefficients for all metrics.
For subsequent experiments, we will therefore focus on this simplified phase-based RT-GCC-NMF formulation using all-ones activation coefficients.\\
\indent Finally, we note that both the number of training and inference iterations offer control over the target fidelity vs. interference suppression trade-off.
While the dictionary pre-learning is performed offline, and thus has no computational effect online, increasing the number of inference iterations comes with additional computational cost at runtime.

\subsubsection{RT-GCC-NMF mask parameter experiments}\label{subsec:experiments-window-parameters}

In Figure \ref{fig:window-function-parameters}, we present the effects on objective speech enhancement quality and intelligibility scores of the RT-GCC-NMF soft-masking function parameters: the TDOA window size $\alpha$, the window floor $\eta$, and the shape parameter $\beta$ from equation \eqref{eq:soft-mask-generation}.
Default settings for these parameters are $\alpha$ = 3/16, $\eta$ = 0, and $\beta=\infty$.
We first note that the TDOA window size $\alpha$ has a drastic effect on the target fidelity vs. interference suppression trade-off, where widening the TDOA window results in reduced interference suppression and higher target fidelity.
Since the target TDOA window width can be controlled online, this provides the most significant control of the trade-off with respect to the parameters presented so far, with no effect on computational requirements.
The highest overall quality scores are achieved for small windows near 1/16 of the TDOA range, while the highest intelligibility scores are achieved for wider windows between 1/4 and 3/8.\\
\indent Effects of the two remaining masking parameters, the window floor $\eta$ and window shape $\beta$, are shown in Figure \ref{fig:window-function-parameters} b) and c) respectively.
Both overall PEASS and BSS Eval scores reach their maxima for $\eta=0$ and $\beta=100$, i.e. for zero window floor and boxcar shaped function, thus reducing to a binary activation coefficient mask.
However, we note peaks in the intelligibility scores for small window floor values near $\eta=0.1$, and window shapes near $\beta=1$, suggesting that while soft-masking may not bring improvement in terms of speech quality, it can improve speech intelligibility.

\begin{table}[t]
\centering
\scriptsize
\caption{Cross-corpora generalization evaluation using different dictionary train set sources and dictionary learning procedures, evaluated on a subset of the SiSEC \emph{dev} dataset. For PEASS and BSS Eval, the overall scores (OPS and SDR) are shown.}
\centering \ra{1.2} 
\tabcolsep=3mm\begin{tabular}{@{}lllllcllll@{}} 
& \multicolumn{1}{c}{PEASS} & \multicolumn{1}{c}{BSS Eval} & \multicolumn{1}{c}{STOI} & \multicolumn{1}{c}{ESTOI} \\ 
\midrule 
SiSEC Copy  & 28.68$\pm$4.77 & 1.46$\pm$1.54 & 0.70$\pm$0.08 & 0.51$\pm$0.09\\ 
CHiME Copy  & 28.24$\pm$5.45 & 1.31$\pm$1.65 & 0.68$\pm$0.09 & 0.50$\pm$0.10\\ 
SiSEC Train  & 32.00$\pm$8.11 & 0.68$\pm$1.10 & 0.76$\pm$0.07 & 0.57$\pm$0.09\\ 
CHiME Train  & 33.94$\pm$7.30 & 1.06$\pm$1.16 & 0.75$\pm$0.07 & 0.56$\pm$0.09\\ 
\bottomrule 
\end{tabular} 
\label{tab:generalizationTable} 
\end{table}

\subsubsection{Effect of input SNR}
To study the effect of input SNR on speech enhancement performance, we recreate all examples from the SiSEC dev dataset at SNRs varying from -40 to 40 dB by rescaling the target speech and noise signals prior to mixing to achieve the desired SNR.
Absolute overall quality and intelligibility measures are presented in Figure \ref{fig:effect-of-snr} a), and relative improvement with respect to the input mixture are shown in Figure \ref{fig:effect-of-snr} b).
Since we found previously that the overall quality and intelligibility scores reach their peaks at different values of the TDOA window width $\alpha$ (Section \ref{subsec:experiments-window-parameters}), we present results here for both a narrow window width, favouring overall enhancement quality metrics, and a wide window width, favouring intelligibility measures.
The wide windows do indeed result in better intelligibility scores over all input SNRs, with improvement in intelligibility scores occurring for input SNRs between -30 and 20 dB.
The narrow window results in significant improvement in SNR for negative input SNRs, with worse reduction in quality for positive input SNRs.
In both cases, the SNR improvement increases monotonically with decreasing input SNR, suggesting that RT-GCC-NMF offers the greatest improvements in the most challenging conditions in terms of SNR.
PEASS scores, on the other hand, suggest that the improvement in speech quality peaks for moderate noise levels, with decreasing improvement for lower and higher input SNRs, with the largest improvement between -10 and 10 dB.
Intelligibility scores also suggest the biggest improvement for input SNRs near 0 dB, bringing improvement from -20 to 10 dB.

\begin{table*}[b]
\caption{Mean PEASS and BSSEval scores $\pm$ standard deviation for different speech enhancement algorithms. RT-GCC-NMF methods include the dictionary pre-learning approach with (a) online localization and (b) offline localization, in addition to (c) offline mixture-learned GCC-NMF for comparison.
Benchmark methods from the SiSEC challenge are presented for comparison, where $^*$ are computed using the subset of examples as reported in \cite{liutkus20172016}, and the ideal binary mask (IBM) is an oracle baseline.} 
\centering \ra{1.2} 
\tabcolsep=0.18cm\begin{tabular}{@{}lllllcllll@{}} 
& \multicolumn{4}{c}{PEASS} & \phantom{abc}& \multicolumn{4}{c}{BSS Eval} \\ 
\cmidrule{2-5} \cmidrule{7-10} 
& \multicolumn{1}{c}{\textbf{OPS}} & \multicolumn{1}{c}{TPS} & \multicolumn{1}{c}{IPS} & \multicolumn{1}{c}{APS} 
&& \multicolumn{1}{c}{\textbf{SDR}} & \multicolumn{1}{c}{ISR} & \multicolumn{1}{c}{SIR} & \multicolumn{1}{c}{SAR} \\ 
RT-GCC-NMF (a)  & 35.87$\pm$3.57 & 41.57$\pm$12.85 & 45.25$\pm$11.75 & 44.23$\pm$5.80 & & 1.40$\pm$1.06 & 4.27$\pm$1.66 & 0.97$\pm$5.54 & 2.95$\pm$0.92\\ 
RT-GCC-NMF (b)  & 37.58$\pm$3.74 & 42.51$\pm$12.33 & 46.32$\pm$10.59 & 45.59$\pm$4.70 & & 1.52$\pm$0.98 & 4.47$\pm$1.72 & 1.10$\pm$5.38 & 2.94$\pm$0.78\\ 
GCC-NMF (c)  & 37.96$\pm$4.52 & 42.39$\pm$12.38 & 45.32$\pm$9.58 & 45.51$\pm$4.57 & & 1.41$\pm$1.18 & 4.83$\pm$1.87 & 0.87$\pm$6.15 & 3.10$\pm$1.16\\ 
 \addlinespace[4pt] \emph{Liu}$^*$\cite{liutkus20172016}  & 14.93$\pm$4.76 & 43.53$\pm$4.45 & 17.13$\pm$1.33 & 69.13$\pm$6.98 & & -7.43$\pm$4.88 & 3.27$\pm$0.97 & -2.00$\pm$5.30 & 14.13$\pm$4.99\\ 
\emph{Duong}$^*$\cite{duong2015speech}  & 16.57$\pm$5.47 & 70.03$\pm$2.93 & 11.53$\pm$5.35 & 73.30$\pm$4.75 & & 7.00$\pm$1.92 & 19.53$\pm$2.53 & 8.57$\pm$2.26 & 12.97$\pm$2.84\\ 
\emph{Rafii} \cite{rafii2013online}  & 30.81$\pm$5.69 & 56.60$\pm$7.74 & 31.53$\pm$9.89 & 55.56$\pm$4.78 & & 5.09$\pm$5.03 & 11.63$\pm$3.39 & 7.98$\pm$6.68 & 8.96$\pm$3.37\\ 
\emph{Magoarou} \cite{arberet2010nonnegative, le2013text}  & 32.66$\pm$7.51 & 66.10$\pm$22.10 & 34.84$\pm$12.71 & 44.13$\pm$11.87 & & 3.67$\pm$5.69 & 17.79$\pm$5.46 & 5.62$\pm$6.36 & 8.77$\pm$4.17\\ 
\emph{Wang} \cite{wang2011region, kayserurl}  & 38.01$\pm$5.79 & 53.91$\pm$9.25 & 54.50$\pm$6.12 & 50.60$\pm$7.02 & & 9.84$\pm$3.09 & 13.54$\pm$4.88 & 19.98$\pm$3.29 & 11.98$\pm$2.76\\ 
 \addlinespace[4pt] \begin{graytext}IBM \cite{ono20152015} \end{graytext} & \begin{graytext}38.37$\pm$9.33 \end{graytext} & \begin{graytext}56.61$\pm$8.96 \end{graytext} & \begin{graytext}73.69$\pm$1.10 \end{graytext} & \begin{graytext}38.43$\pm$10.54 \end{graytext} & & \begin{graytext}14.42$\pm$2.81 \end{graytext} & \begin{graytext}26.08$\pm$3.91 \end{graytext} & \begin{graytext}23.89$\pm$2.90 \end{graytext} & \begin{graytext}15.20$\pm$2.98 \end{graytext} \\ 
\bottomrule 
\end{tabular} 
\label{tab:algorithmComparisonTable} 
\end{table*}

\subsubsection{Comparison between approaches}\label{sec:Experiments-Online}
In Table \ref{tab:gccNMFResultsTable}, we compare the performance of several elements of RT-GCC-NMF including dictionary pre-learning vs. mixture learning, activation coefficient inference vs. all-ones coefficients, moderate vs. large NMF dictionaries, and offline vs. online localization, for the static speaker case.
We first note that replacing activation coefficients with the all-ones vector, as described in Section \ref{subsec:coefficient-inference}, results in a substantial increase in performance across PEASS and intelligibility scores in all cases, despite decreasing the SNR-based BSS Eval score.
Performing target localization online, as described in Section \ref{subsec:online-localization}, results in somewhat decreased in performance in almost all cases.
A more robust approach to target localization than the simple accumulated GCC-PHAT is therefore likely necessary in practice.
Pre-learning the NMF dictionary results in similar performance to the previously presented offline mixture-learned approach.
As the dictionary is pre-learned using a different dataset (CHiME) than used for testing (SiSEC), this approach generalizes to unseen speakers, acoustic conditions, and recording setups.
Finally, we note that the all-ones activation coefficients approach also results in increased performance for the offline mixture-trained GCC-NMF approach we had presented previously \cite{wood2017blind}, exhibiting significantly improved performance in terms of PEASS and intelligibility measures.
Offline GCC-NMF can therefore also benefit from this finding.\\
\indent To further study the ability of the proposed RT-GCC-NMF to generalize across datasets, we proceed to determine whether its performance is affected when the inference dataset is different from the dataset used to train the NMF dictionary.
The SiSEC \emph{dev} set is first split into train and evaluation subsets where inference is performed on the second half of each example, while the first half is used for training.
This split guarantees that the same speakers and noise types are present in both train and evaluation subsets, with different instances occurring in both.
The \emph{within-corpus} dictionary is then trained using the first half of the SiSEC dataset, while the \emph{cross-corups} dictionary is trained on CHiME.
In Table \ref{tab:generalizationTable}, we present results for both the standard dictionary training method as well as the \emph{copy-to-train} dictionary construction method presented in \cite{king2010single}, where randomly-selected frames from the training data are used as the dictionary atoms.
We first note that there is no significant difference in performance between the \emph{within-corpus} dictionary and the \emph{cross-corpus} dictionary, for both dictionary construction methods.
Second, we note improved performance when using train method over the copy method for both the \emph{within-corpus} and \emph{cross-corpus} dictionaries.
This may be due to NMF learning a more fine-grained \emph{parts-based} representation, with the resulting mask constructed by associating more low-level parts instead of complete spectra as with the \emph{copy-to-train} approach.\\
\indent In Table \ref{tab:algorithmComparisonTable}, we compare RT-GCC-NMF with other algorithms from the 2013, 2015, and 2016 SiSEC separation challenges \cite{ono2013, ono20152015, liutkus20172016}.
We use the SiSEC \emph{dev} set for comparison here as the \emph{test} set was evaluated by the challenge organizers, and isolated speech and noise signals necessary for evaluation are not publicly available.We therefore use the default parameters defined in Section \ref{sec:experimental-setup} and do not optimize the parameters using the analysis presented in Section \ref{subsec:experiments-online}.
We present three variants in Table \ref{tab:algorithmComparisonTable}: offline GCC-NMF with mixture-learned dictionary, and pre-learned dictionary RT-GCC-NMF with and without online localization, each using the phase-based, all-ones activation coefficient technique.
For the offline GCC-NMF, the NMF dictionary is trained directly on the SiSEC mixture signals, while for the RT-GCC-NMF, the dictionary is trained using isolated speech and noise signals from the CHiME dataset as described in Section \ref{sec:experimental-setup}.
All approaches outperform all but one of the previous methods, most of which rely on supervised learning or are unsuitable in online settings.
The method that outperforms RT-GCC-NMF is a frequency-domain blind source separation technique using a region growing permutation alignment approach \cite{wang2011region}.
While the authors show that it has the potential to run in real-time, a real-time implementation has not yet been presented to the best of our knowledge.
These results demonstrate that RT-GCC-NMF holds significant potential for future research and applications, especially given that it remains purely unsupervised, conceptually simple, easy to implement, and generalizes across speakers, acoustic environments, and recording setups.

\begin{figure}[t!]
\centering
\includegraphics{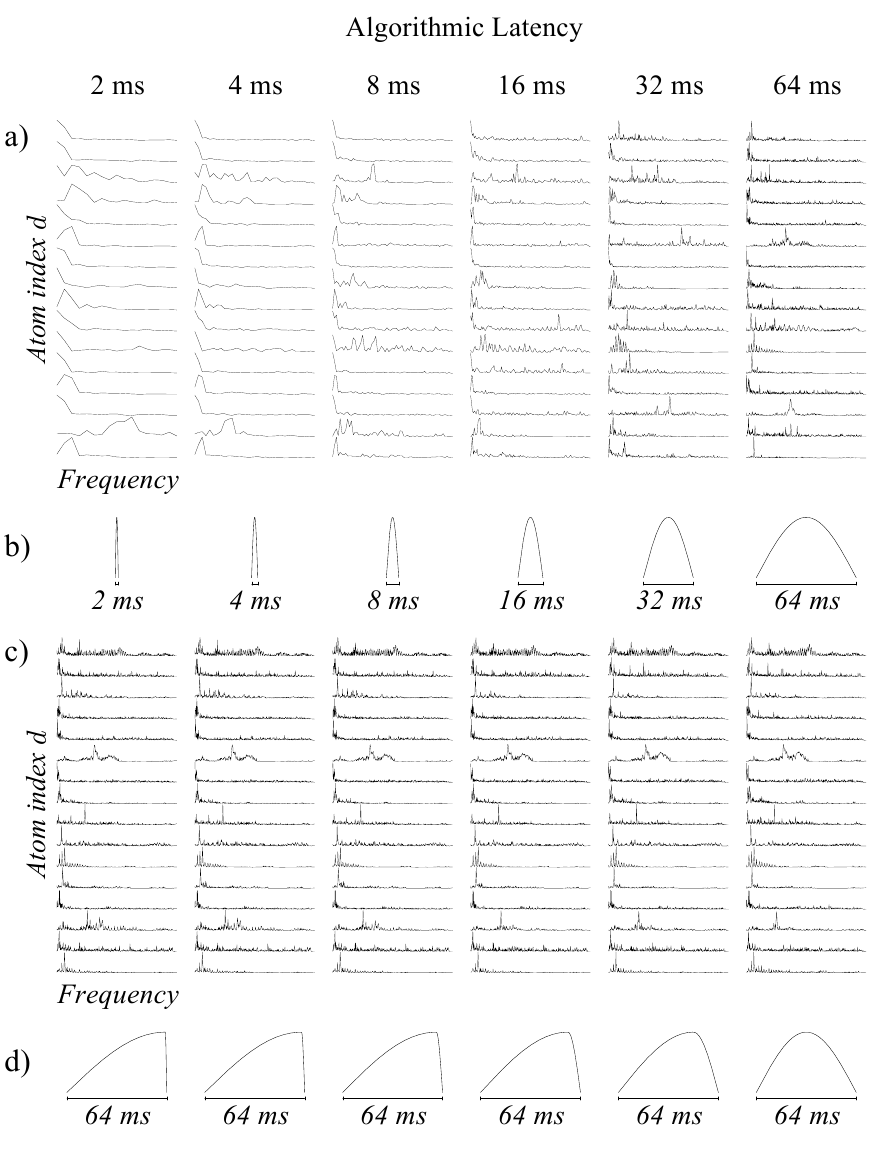}
\caption{Example NMF dictionary atoms and corresponding STFT analysis windows for varying algorithmic latencies for the symmetric windowing (a, b) and asymmetric windowing (c, d) latency reduction strategies.
For each algorithmic latency value, a subset of 16 randomly chosen atoms are shown from a total of 1024.
For symmetric windowing, the analysis window size decreases with algorithmic latency, while for asymmetric windowing, the analysis window size remains constant at 64 ms, while its shape changes as a function of latency.}
\label{fig:dictionaries-varying-window-size}
\end{figure}

\subsection{Low latency RT-GCC-NMF experiments}\label{sec:experiments-low-latency}
In this section, we evaluate the low latency version of RT-GCC-NMF.
We begin by comparing the effect of latency reduction of the symmetric and asymmetric windowing methods on the learned NMF dictionary atoms, followed by the effect of both approaches on the objective speech enhancement quality and intelligibility measures.
We then study the empirical processing time requirements of low latency RT-GCC-NMF for a variety of hardware platforms to determine the conditions under which the proposed system may run in real-time on currently available hardware platforms.

\subsubsection{Asymmetric windowing and NMF dictionary atoms}\label{subsec:enhancement-quality-experiments}
In Figure \ref{fig:dictionaries-varying-window-size} a), we depict example NMF dictionary atoms learned using the \emph{symmetric} STFT windowing method, for varying algorithmic latencies.
We note that as the window size is decreased, atoms become increasingly wideband, and the spectral details captured with longer duration windows are lost.
Contrary to the traditional windowing approach, \emph{asymmetric} windowing allows us to retain the long-duration analysis windows while decreasing the synthesis window size.
As the synthesis window size $2M$ is reduced, the analysis window size remains fixed at the frame size $N$, with its shape increasingly weighted towards the future.
Example NMF dictionary atoms learned using the asymmetric windowing approach with varying algorithmic latencies are shown in Figure \ref{fig:dictionaries-varying-window-size} c).
We note that the learned NMF dictionary atoms retain spectral detail independent of synthesis window size.
Since identical training data and random seed is used in all cases, the resulting atoms remain very similar across all algorithmic latencies, with subtle differences in the learned NMF dictionary atoms resulting from the different analysis window shapes.
\begin{figure}[h]
\includegraphics{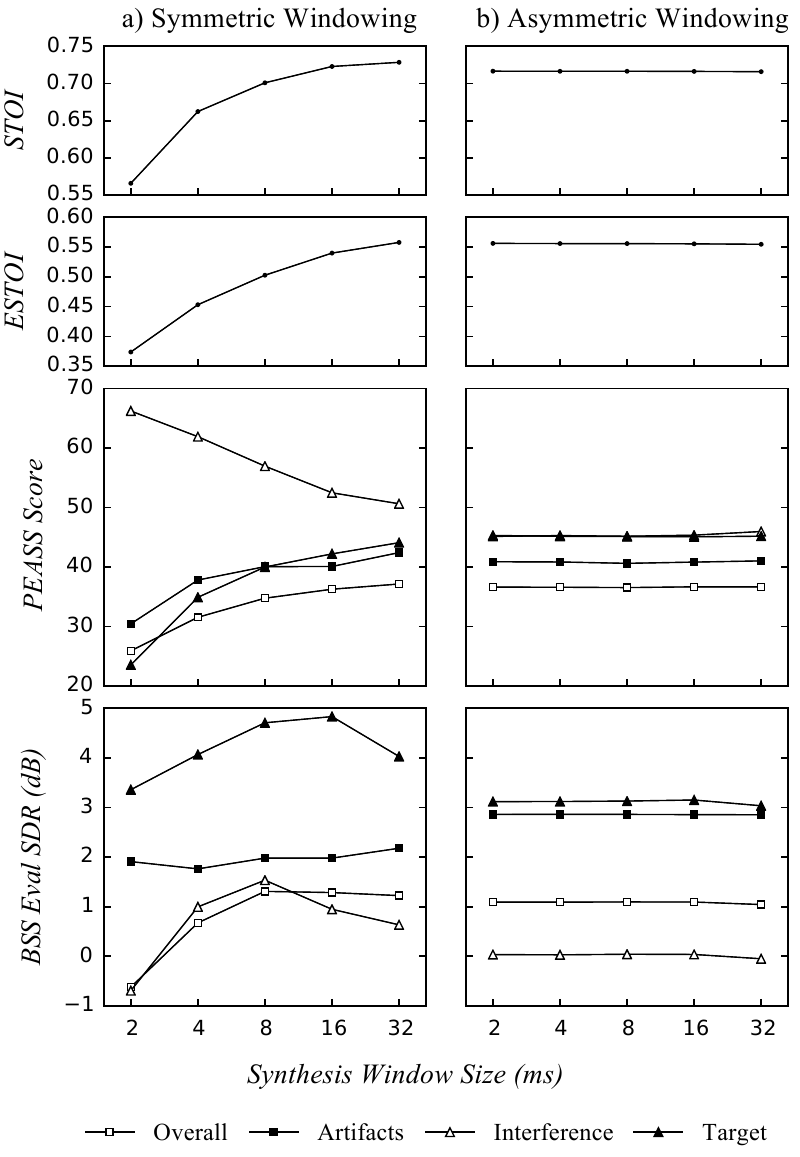}
\caption{Effect of STFT synthesis window size on RT-GCC-NMF speech enhancement performance for a) symmetric windowing and b) asymmetric windowing with a fixed analysis window of 64 ms.
PEASS and BSS Eval scores are as described in Figure \ref{fig:train-dictionary-size}.}
\label{fig:effect-of-latency-reduction}
\end{figure}

\subsubsection{Asymmetric windows and speech enhancement quality}\label{subsec:enhancement-quality-experiments}
In Figure \ref{fig:effect-of-latency-reduction} a), we present the objective speech enhancement quality and intelligibility measures as a function of algorithmic latency for the symmetric windowing case.
We note that both the overall quality scores as well as intelligibility scores decrease with decreasing window size, with a significant drop in PEASS overall performance for window sizes less than 8 ms.
This is likely due to a decreased separability of speech and noise sources with the wideband NMF dictionary atoms shown in Figure \ref{fig:dictionaries-varying-window-size} a), resulting in decreased quality of the resulting RT-GCC-NMF speech enhancement.
We also note a significant trade-off between interference suppression and both target fidelity and artifact PEASS scores, where smaller window sizes result in increased interference suppression at the cost of significant artifacts and poorer target fidelity.\\
\indent In Figure \ref{fig:effect-of-latency-reduction} b), we present the effect of latency for the asymmetric windowing approach in the same conditions as above.
The analysis window here is kept fixed at 1024 samples at 16 kHz (64 ms), while the synthesis window size is varied from 512 to 32 samples (32 to 2 ms), with an overlap of 75\% of the synthesis window used in each case.
We note that all scores remain relatively constant for varying synthesis window size, even for latencies as low as 2 ms.
These results demonstrate that the proposed asymmetric windowing approach is a viable solution to reduce the latency of RT-GCC-NMF to values well below the threshold required for hearing devices, while maintaining the enhancement quality of the higher latency symmetric windowing approach.
\begin{table*}[t]


\caption{Hardware platform specifications for experiments in Section \ref{subsec:latency-processing-time}. $^\dagger$Hosted on the same platform.}
\centering \ra{1.2} 

\newcommand{\na}{-}

\tabcolsep=0.16cm\begin{tabular}{@{}lcccccccccc@{}} 
& \multicolumn{4}{c}{CPU} & \phantom{abc} & \multicolumn{3}{c}{GPU} & \phantom{a} & \multicolumn{1}{c}{Power Consumption}\\ 
\cmidrule{2-5} \cmidrule{7-9} \cmidrule{11-11}
& \multicolumn{1}{c}{Type} & \multicolumn{1}{c}{Cores} & \multicolumn{1}{c}{Clock Speed} & \multicolumn{1}{c}{RAM} 
&& \multicolumn{1}{c}{Type} & \multicolumn{1}{c}{Cores} & \multicolumn{1}{c}{RAM} 
&& \multicolumn{1}{c}{Theoretical Max.} \\ 
Raspberry Pi 3 & ARM A53 & 4 & 1.2 GHz & 1 GB && \na & \na & \na & & 4 W\\ 
Jetson TX1 GPU & ARM A57 & 4 & 1.9 GHz & 4 GB && NVIDIA Maxwell & 256 & 4 GB & & 15 W\\
MacBook Pro & Intel Core 2 Duo & 2 & 2.4 GHz & 4 GB && \na & \na & \na & & 55 W\\ 
Xeon$^\dagger$, Tesla$^\dagger$ & Intel Xeon E5-1620v3 & 4 & 3.5 GHz & 16 GB && NVIDIA Tesla K40C & 2880 & 12 GB & & 275, 500 W\\ 
\bottomrule 
\end{tabular} 
\label{tab:hardwareSpecsTable}


\end{table*}

\subsubsection{Latency and RT-GCC-NMF processing time}\label{subsec:latency-processing-time}
\begin{figure}[h]
\centering
\includegraphics{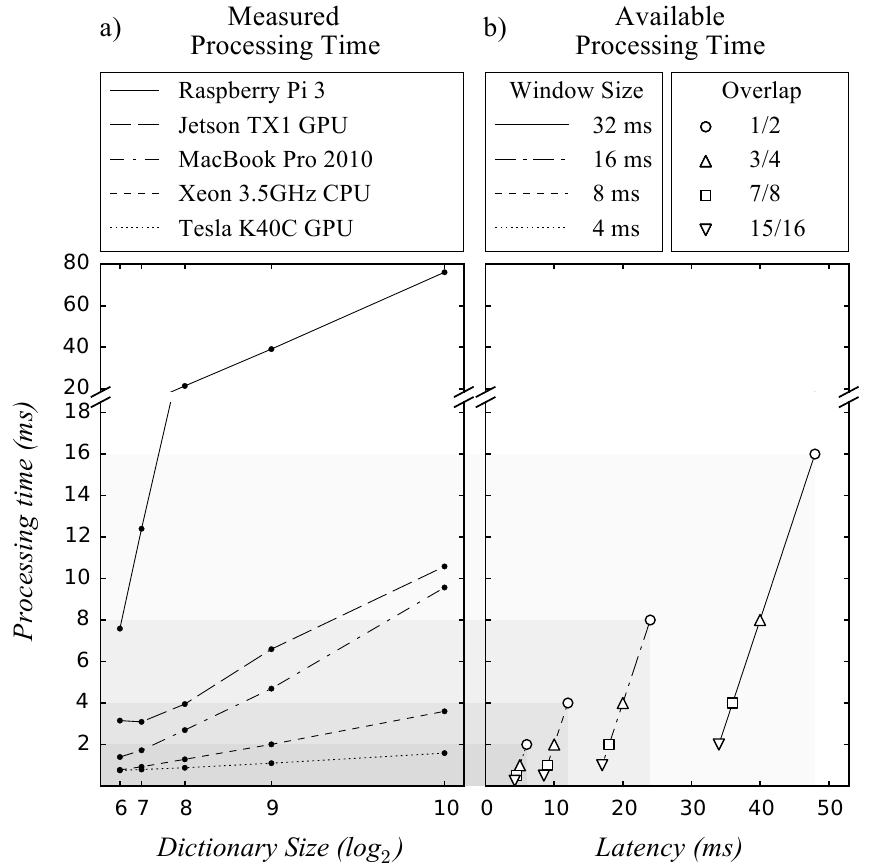}
\caption{RT-GCC-NMF computational requirements with the asymmetric STFT windowing technique, with a) Effect of dictionary size on RT-GCC-NMF mean empirical processing time for a single frame on various hardware platforms given an analysis window size of 64 ms, and b) available processing time for a single frame, given the asymmetric STFT windowing approach, presented for varying synthesis window size and overlap, with the resulting latency as the horizontal axis.}
\label{fig:processing-time}
\end{figure}

We now proceed to study the computational requirements of the RT-GCC-NMF speech enhancement algorithm with asymmetric windowing to determine the conditions under which it may be run in real-time.
As we saw in Section \ref{sec:asymmetric-stft-windowing}, the latency of the asymmetric STFT process in a real-time system is equal to the duration of the synthesis window plus the frame advance.
For speech enhancement to be performed in real-time, the system must then process a single frame within the time of a single frame advance.
This processing time includes the windowing processes, the forward FFT, the RT-GCC-NMF speech enhancement processing itself, the inverse FFT, and the OLA summation.
To evaluate processing time empirically, we use the RT-GCC-NMF implementation written in Python with the Theano optimizing compiler \cite{theanourl} as described in Section \ref{sec:realtime-implementation}.\\
\indent In Figure \ref{fig:processing-time} a), we present the average measured processing time of a \emph{single frame}, as a function of the NMF dictionary size, for a variety of hardware platforms (see Table \ref{tab:hardwareSpecsTable} for specifications).
We note that the processing time increases approximately linearly with dictionary size, with the slope varying between hardware platforms.
On all systems presented, processing times less than 8 ms are possible, provided a small enough dictionary is used.
Since speech enhancement performance decreases smoothly with decreasing dictionary size as we showed in Section \ref{subsec:dictionary-pretraining}, Figure \ref{fig:train-dictionary-size} a), RT-GCC-NMF can be easily adapted to a range of hardware platforms, with performance determined by available computational power.
Finally, we note that since the asymmetric windowing approach only affects the shape of the analysis and synthesis windows, it has no effect on the computational demands of the algorithm for a single frame.
Exactly the same operations are performed in both cases, where only the sampled values of the window functions differ.
However, asymmetric windowing does result in more \emph{overall} computation as more frames need to be processed due to the decreased frame advance required for short synthesis windows.\\
\indent In Figure \ref{fig:processing-time} b), we depict the relationship between the inherent STFT latency and available processing time for a single frame, as a function of synthesis window size and overlap.
The available processing time is determined analytically for a given window size and overlap, and does not reflect empirical, measured values.
Instead, this Figure depicts the maximum amount of time that a given hardware device can take to process a single frame in order to operate in real-time.
We note that decreasing either the synthesis window size or the frame advance decreases the system latency at a cost of decreased available processing time.
We may combine this information with Figure \ref{fig:processing-time} a) to determine, for a given hardware platform and dictionary size, the available synthesis window size and overlap values (and resulting latencies), for the system to run in real-time.\\
\indent Experiments showed that enhancement performance was unaffected by window overlap for values greater than 75\%, with a small decrease in performance with an overlap of 50\%.
Focusing on the fastest possible performance here, we consider a window overlap of 50\% in the following.
All systems prove fast enough for a synthesis window size of 16 ms with 50\% overlap and a dictionary size of 64, resulting in a latency of 24 ms.
All systems except the Raspberry Pi may achieve 12 ms latency for small to moderate dictionary sizes, with a window size of 8 ms and 50\% overlap.
The fastest system (Tesla K40 GPU) can achieve 6 ms latency for dictionaries at least as large as 1024 atoms.
It is therefore possible to achieve latencies suitable for hearing assistive devices on modern embedded and desktop computing platforms using speech enhancement with RT-GCC-NMF with the asymmetric windowing latency reduction technique.
Future work will involve additional implementation optimizations in an effort to run RT-GCC-NMF on lower-power devices including digital signal processors (DSP) that are better-suited to wearable real-world use for hearing assistive applications than the platforms presented here.

\section{Real-time implementation}\label{sec:realtime-implementation}
RT-GCC-NMF was written in Python, using the Theano optimizing compiler \cite{theanourl}, with an interactive interface using PyQt \cite{pyqturl} and pyqtgraph \cite{pyqtgraphurl}.
The graphical user interface allows users to manipulate both NMF and masking function parameters in real-time, such that their effects on subjective enhancement quality and intelligibility can be studied interactively.
The user may also specify the target TDOA location manually or enable automatic localization where the effect of the sliding window width can be manipulated in real-time.
Examples of both static and moving speakers are provided.
The software has been tested on a wide range of hardware platforms where performance can be made to degrade smoothly with decreasing computational power by using smaller pre-learned dictionaries, as we showed in Figure \ref{fig:train-dictionary-size}.
Source code for RT-GCC-NMF and iPython notebook demonstrations are made available as open source at \mbox{\url{https://www.github.com/seanwood/gcc-nmf}}.

\section{Conclusion}\label{sec:conclusion}
We have presented a low latency speech enhancement algorithm called RT-GCC-NMF, and studied its performance on stereo mixtures of speech and real-world noise.
We showed that by pre-learning the NMF dictionary in a purely unsupervised fashion on a different dataset than used at runtime, RT-GCC-NMF generalizes to new speakers, acoustic environments, and recording setups.
This approach is therefore flexible in terms of training data, where training is not bound to a speaker-specific dataset or to test data from a limited number of speakers.
Also, only a very small amount of unlabelled training data is required, on the order of one thousand 64 ms frames, significantly less than the hours or days of labeled training data required by supervised deep learning approaches to speech enhancement.
RT-GCC-NMF holds significant potential for future research as it is conceptually simple, easy to implement, purely unsupervised, and generalizes to unseen datasets.\\
\indent A phase-based version of RT-GCC-NMF was developed that bypasses NMF activation coefficient inference, such that only the pre-learned dictionary and input phase differences are required at runtime.
This method outperformed all but one algorithm previously submitted to the SiSEC speech enhancement challenge, comparing favourably to the state-of-the-art and the ideal binary mask (IBM) baseline.
This approach simultaneously improved both objective speech quality and intelligibility metrics over a wide range of input SNRs.
An interesting direction for future work would be to combine phase-based RT-GCC-NMF with other phase-aware approaches to further improve estimation of the Wiener filter, or to estimate the phase spectra itself, as it has been shown that it is possible to outperform the IBM baseline by enhancing the spectral phase prior to reconstruction \cite{mayer2015improved,kulmer2016single}.\\
\indent We presented a soft NMF activation coefficient masking alternative to the binary coefficient masking function, and showed that the trade-off between interference suppression and target fidelity can be controlled frame-by-frame via the target TDOA window width, with no effect on computational cost.
The trade-off between speech quality and speech intelligibility can also be controlled on a frame-by-frame basis via the masking function parameters.
In the context of hearing assistive devices, users could therefore be given control of the soft-masking parameters, such that they could be modified depending on their needs for intelligibility, quality, or interference suppression for a given situation.\\
\indent We drastically reduced the inherent algorithmic latency of RT-GCC-NMF by incorporating an asymmetric STFT windowing scheme proposed by Mauler and Martin \cite{mauler2007low}.
Objective speech enhancement quality and intelligibility metrics were shown to remain unaffected over latencies from 32 ms to 2 ms, where the NMF dictionary atoms adapted to the changing analysis window shapes.
This latency reduction falls well within the range of tolerable latencies for hearing aids, i.e. 10 ms or less for the general case.\\
\indent Finally, we developed an open source implementation of RT-GCC-NMF, allowing subjective analysis of the effects of various system parameters to be studied interactively in real-time via a graphical user interface.
We showed that latencies suitable for use in hearing assistive applications were achievable on a variety of hardware platforms ranging from desktop PCs to low-cost embedded system on a chips (SoCs).
Future work will include further algorithmic and memory optimizations to run RT-GCC-NMF on lower-power devices suitable for real-world hearing assistive applications.

\section*{Acknowledgments}
The authors would like to thank ACELP/CEGI, NSERC, and FQRNT (CHIST-ERA, IGLU) for funding our research, as well as the developers of the open-source PEASS, BSSEval libraries.
We would also like to thank William E. Audette and Simon Brodeur for inspiring discussions during the development of this work.
The research was enabled in part by support provided by Calcul Quebec (www.calculquebec.ca) and Compute Canada (www.computecanada.ca).

\bibliographystyle{IEEEtran}
\bibliography{IEEEabrv,RTGCCNMF}

\begin{IEEEbiography}[{\includegraphics[width=1in,height=1.25in,clip,keepaspectratio]{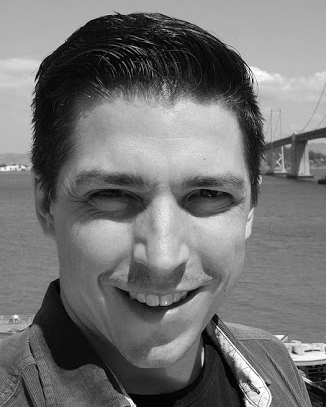}}]
{Sean U. N. Wood} received the B.A.Sc. degree in engineering science from the University of Toronto, Toronto, ON, Canada in 2004, followed by the M.Sc. degree in computer science at the Montreal Institute for Learning Algorithms (MILA) from the Universit\'{e} de Montr\'{e}al, Montreal, QC, Canada, in 2010, and the Ph.D. degree in electrical engineering at the Computational Neuroscience and Intelligent Signal Processing Research (NECOTIS) group at the Universit\'{e} de Sherbrooke, Sherbrooke, QC, Canada in 2017.
He is currently a Postdoctoral Researcher with the Signal Processing and Speech Communication (SPSC) Laboratory at the Graz University of Technology in Graz, Austria.
His research combines machine learning, multi-channel signal processing, and computational neuroscience.
He is particularly interested in unsupervised learning algorithms, real-time systems, and applications in speech and biomedical signal processing.
\end{IEEEbiography}
\vfill
\newpage
\begin{IEEEbiography}[{\includegraphics[width=1in,height=1.25in,clip,keepaspectratio]{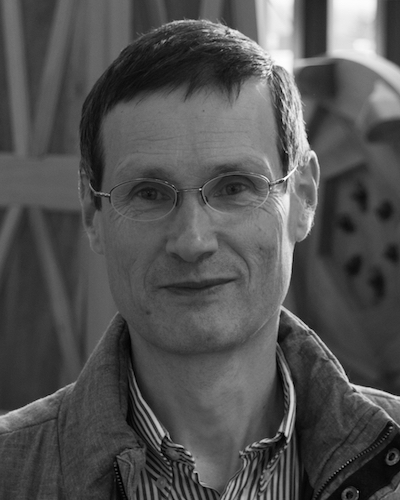}}]
{Jean Rouat}(S'82-M'88-SM'06) received the M.Sc. degree in physics from the Universit\'e de Bretagne, Brest, France, in 1981, the E. \& E. M.Sc.A. degree in speech coding and speech recognition from the Universit\'e de Sherbrooke, QC, Canada, in 1984, and the E. \& E. Ph.D. degree in cognitive and statistical speech recognition jointly with the Universit\'e de Sherbrooke and McGill University, Montreal, QC, in 1988.
He held a post-doctoral position in psychoacoustics with the MRC, App. Psych. Unit, Cambridge, U.K., and in electrophysiology with the Institute of Physiology, Lausanne, Switzerland.
He was an Adjunct Professor with the Department of Biological Sciences, Universit\'e de Montréal from 2007 to 2018.
He is currently with the Universit\'e de Sherbrooke, where he founded the Computational Neuroscience and Intelligent Signal Processing Research Group, and is Full Member of the Centre for Interdisciplinary Research in Music Media and Technology, Schulich School of Music, McGill University, Montreal.
His translational research links neuroscience and engineering for the creation of new technologies and applications with the integration of a better understanding and integration of multimodal representations (vision and audition).
Information hiding in multimedia signals, development of hardware low power consumption neural processing units for a sustainable development, interactions with artists for multimedia and musical creations are examples of transfers that he leads based on the knowledge he gains from neuroscience and his knowledge of visual and auditory systems.
He is leading machine learning funded projects to develop sensory substitution and intelligent devices.
He is also the coordinator of the interdisciplinary IGLU CHIST-ERA European Consortium (IGLU - Interactive Grounded Language Understanding) for the development of an intelligent agent that learns through multimodal grounded interactions.
\end{IEEEbiography}

\end{document}